\documentclass[a4paper,11pt]{article}
\pdfoutput=1
\usepackage[english]{babel}
\usepackage{jheppub}
\usepackage{amsmath,amssymb,graphicx,textcomp,enumerate,xspace}
\usepackage{soul} 

\usepackage[utf8]{inputenc}

\newcommand\RES{{\tt POWHEG~\!BOX~\!RES}}

\newcommand{\mathd}{\mathrm{d}}

\newcommand{\tmop}[1]{\ensuremath{\rm #1}}

\newcommand\sss{\mathchoice%
{\displaystyle}%
{\scriptstyle}%
{\scriptscriptstyle}%
{\scriptscriptstyle}%
}
\newcommand{\pt}{p_{\sss\rm T}}
\newcommand{\kt}{{\ensuremath{k_{\rm\sss T}}\xspace}}

\newcommand{\ktmin}{\ensuremath{k_{\sss\rm T}^{\rm \sss min}}}

\newcommand{\muF}{\ensuremath{\mu_{\mathrm{F}}}}
\newcommand{\muR}{\ensuremath{\mu_{\mathrm{R}}}}

\def\beq{\begin{equation}}
\def\beqn{\begin{eqnarray}}
\def\eeq{\end{equation}}
\def\eeqn{\end{eqnarray}}

\def\lq{\left[} 
\def\rq{\right]} 
\def\rg{\right\}} 
\def\lg{\left\{} 
\def\({\left(} 
\def\){\right)} 
 
\newcommand\nf{n_{\rm f}}

\newcommand\as{\alpha_{\sss\rm S}}

\newcommand\CF{C_{\sss\rm F}}
\newcommand\CA{C_{\sss\rm A}}

\newcommand\TR{T_{\sss\rm R}}

\newcommand\muf{\mu_{\sss\rm F}}
\newcommand\mur{\mu_{\sss\rm R}}

\newcount\minutes 
\newcount\scratch 
\def\timestamp{%
\scratch=\time 
\divide\scratch by 60 
\edef\hours{\the\scratch} 
\multiply\scratch by 60 
\minutes=\time 
\advance\minutes by -\scratch 
---$\,$\hours:\null 
\ifnum\minutes< 10 0\fi 
\the\minutes} 
\newcommand\aaj{\gamma\gamma j}

\newcommand{\MSB}{\ensuremath{\overline{\rm MS}}}

\def\beeq{\begin{eqnarray}} 
\def\eeeq{\end{eqnarray}} 
\def\to{\rightarrow}

\newcommand\Phirad{\Phi_{\sss {\rm rad}}}

\newcommand\HCCF{\lq \HCCF H^{F} C_1 C_2 \rq_{c\bar{c};\,ab}}

\newcommand\Lum{{\cal L}}
\newcommand\ord[1]{\mathcal{O}\!\(#1\)}

\newcommand\pti{p_{\sss \rm T}^{\sss i}}

\newcommand\dontshow[1]{}
\newcommand{\MiNNLOPS}{{\sc\small MiNNLO\textsubscript{PS}}}

\newcommand{\MiNLO}{{\sc\small MiNLO}}

\newcommand{\pwg}{{\sc\small Powheg}}
\newcommand{\Powheg}{{\sc\small Powheg}}

\newcommand{\pwgboxres}{{\sc\small Powheg Box Res}}
\newcommand{\pythia}{{\sc\small Pythia}}

\newcommand{\Radish}{{\sc\small RadISH}}

\newcommand{\rivet}{{\sc\small {Rivet}}} 
\newcommand{\Matrix}{{\sc\small {Matrix}}} 
\newcommand{\PowhegBox}{{\sc\small Powheg Box}} 
\newcommand{\LHAPDF}{{\sc\small Lhapdf}} 
\newcommand{\OpenLoopsTwo}{{\sc\small OpenLoops2}} 
\newcommand{\PythiaEight}{{\sc\small Pythia8}} 

\newcommand{\MAX}{{\sss\rm max}} 
\newcommand{\MIN}{{\sss\rm min}} 
\newcommand{\rad}{{\sss\rm rad}} 
\newcommand{\tr}{{\sss\rm T}} 

\newcommand{\ETmax}{{E_\tr^\MAX}}
\newcommand{\mgamgam}{{m_{\gamma\gamma}}}
\newcommand{\mgamgammin}{{m_{\gamma\gamma}^\MIN}}

\newcommand{\pTgamone}{{p_{\tr\gamma_1}}}
\newcommand{\pTgamonemin}{{p_{\tr\gamma_1}^\MIN}}
\newcommand{\pTgamtwo}{{p_{\tr\gamma_2}}}
\newcommand{\pTgamtwomin}{{p_{\tr\gamma_2}^\MIN}}
\newcommand{\Rcone}{{R_{\sss\rm c}}}

\newcommand\Phin{{ \Phi}_n}
\newcommand\Phinpo{{\Phi}_{n+1}}

\newcommand\alr{{\alpha_{\rm\sss r}}}

\newcommand\PhiB{\Phi_{\sss \gamma\gamma}}
\newcommand\PhiBj{\Phi_{\sss \gamma\gamma j}}
\newcommand\PhiBjj{\Phi_{\sss \gamma\gamma jj}}

\newcommand{\Fcorr}{F^{\tmop{corr}}\!\(\PhiBj\)}
\newcommand{\abar}{\frac{\as}{2\pi}}
\newcommand{\abarmu}[1]{\frac{\as(#1)}{2\pi}}

\newcommand{\sigmas}{\sigma_{\sss \! s}}
\newcommand{\sigmaf}{\sigma_{\sss \! f}}

\newcommand{\pgo}{p_{\gamma_1}}
\newcommand{\pgt}{p_{\gamma_2}}

\newcommand{\ppgo}{\bar{p}'_{\gamma_1}}
\newcommand{\ppgt}{\bar{p}'_{\gamma_2}}

\newcommand{\thb}{\bar{\theta}}
\newcommand{\phib}{\bar{\varphi}}
\newcommand{\pj}{p_{\sss j}}
\newcommand{\phij}{\phi_{\sss j}}
\newcommand{\thj}{\theta_{\sss j}}

\newcommand{\thp}{\theta'}
\newcommand{\phip}{\varphi'}
\newcommand{\ggj}{\gamma\gamma j}

\newcommand{\noFOatQ}{{\tt no\,FOatQ}}
\newcommand{\FOatQ}{{\tt FOatQ}}

\newcommand\ptAA{p_{{\sss\rm T}\gamma\gamma}}
\newcommand{\MatrixRadish}{{\sc\small {Matrix+RadISH}}}

\newcommand\SB{{S_{\sss\rm B}}}
\newcommand\SR{{S_{\sss\rm R}}}

\title{NNLO+PS Monte Carlo simulation of photon pair production with {\sc
    MiNNLO\textsubscript{PS}}}

\author[a,b]{Alessandro Gavardi,}
\author[a,b]{Carlo Oleari,}
\author[a,b,1]{Emanuele Re\note{On leave of absence from LAPTh, Universit\'e Grenoble Alpes,
 Universit\'e Savoie Mont Blanc, CNRS, F-74940 Annecy, France.}}

\emailAdd{a.gavardi@campus.unimib.it}
\emailAdd{carlo.oleari@mib.infn.it}
\emailAdd{emanuele.re@mib.infn.it}

\affiliation[a] {Universit\`a degli Studi di Milano\,-\,Bicocca,
  Piazza della Scienza 3, 20126 Milano, Italy}
\affiliation[b] {INFN, Sezione di Milano\,-\,Bicocca,
  Piazza della Scienza 3, 20126 Milano, Italy}

\preprint{LAPTH-010/22}

\abstract{
  We present a NNLO QCD accurate event generator for direct photon pair
  production at hadron colliders, based on the \MiNNLOPS{} formalism, within
  the \pwgboxres{} framework.
  Despite the presence of the photons requires the use of isolation criteria,
  our generator is built such that no technical cuts are needed at any stage
  of the event generation.  Therefore, our predictions can be used to
  simulate kinematic distributions with arbitrary fiducial cuts.
  Furthermore, we describe a few modifications of the \MiNNLOPS{} formalism
  in order to allow for a setting of the renormalization and factorization
  scales more similar to that of a fixed-order computation, thus reducing the
  numerical impact of higher-order terms beyond the nominal accuracy.
  Finally, we show several phenomenological distributions of physical
  interest obtained by showering the generated events with \PythiaEight{},
  and we compare them with the 13~TeV data from the ATLAS Collaboration.
      
}

\keywords{NLO and NNLO computations, perturbative QCD, photon production, resummation.  
}

\begin{document}

\maketitle

\section{Introduction}
\label{sec:intro}

The production of two isolated photons at hadron colliders, henceforth
denoted diphoton production and abbreviated as $pp\to\gamma\gamma$, is one
among the most relevant Standard Model~(SM) processes, due, on one side, to
the high production rate, and, on the other, to the relative cleanliness
of the experimental final state.
Furthermore, diphoton production is the dominant background for studies
involving Higgs boson production decaying into a photon pair, and, notably,
it was one of the dominant backgrounds for the Higgs boson
discovery~\cite{ATLAS:2012yve, CMS:2012qbp} at the Large Hadron
Collider~(LHC).  In addition, it is a background for searches for heavy
neutral resonances that can arise in a variety of beyond the Standard Model
scenarios (see e.g.~\cite{Mrenna:2000qh,Han:1998sg, Giudice:1998bp,
  Giudice:2017fmj}), and for searches for extra dimensions or supersymmetry,
and a possible channel for their detection~\cite{ATLAS:2017ayi, CMS:2018dqv}.

Within the SM, photon pairs can be produced by means of several mechanisms,
that make this otherwise simple process particularly challenging to study. In
this work we restrict to prompt photon production,\footnote{One refers to
non-prompt photons to denote photons produced from hadron decays.} where
photons are produced in the hard process. Within this category, one can
further distinguish two components: one arising from direct photon
production and the other from single and double fragmentation. In the former
case, the photons are directly produced in the hard scattering, whereas, in
the latter case, one or both photons are produced through the fragmentation
of jets. The latter mechanism can be suppressed by imposing the isolation of
the photons from the hadronic activity through a fixed or a smooth
cone algorithm~\cite{Frixione:1998jh}.
In spite of the fact that the fixed-cone
isolation criterion is simpler to apply (also at the experimental level),
only the smooth-cone isolation is suitable for theoretical calculations, as
it allows to completely remove the fragmentation component in an
infrared-safe way. 

Theoretical predictions for diphoton direct production at
next-to-leading-order~(NLO) QCD accuracy, that consistently include also the
fragmentation component, were obtained for the first time in
ref.~\cite{Binoth:1999qq}, and made publicly available in the {\tt DIPHOX}
software.
Next-to-next-to-leading-order~(NNLO) QCD corrections to $pp\to\gamma\gamma$
were obtained in
refs.~\cite{Catani:2011qz,Campbell:2016yrh,Catani:2018krb,Gehrmann:2020oec},
whereas NLO electroweak corrections have been studied in
ref.~\cite{Bierweiler:2013dja}. More recently, even the three-loop
$q\bar{q}\to \gamma\gamma$ amplitudes have been
computed~\cite{Caola:2020dfu}. In order to include all the effects up to
$\as^2$ accuracy, the gluon-induced production of a photon pair through a
closed quark loop has to be taken into account as well: such contribution,
that is particularly sizable due to the gluon luminosity, was first computed
in ref.~\cite{Dicus:1987fk}, and it is by now known at the next order, both
for massless~\cite{Bern:2002jx} and for massive quark
loops~\cite{Maltoni:2018zvp, Chen:2019fla}.  Substantial progress has also
been made for the computation of a photon pair in association with one or
more jets: predictions for diphoton in association with up to three jets,
including NLO QCD corrections, have been obtained in
refs.~\cite{DelDuca:2003uz,
  Gehrmann:2013aga, Gehrmann:2013bga, Badger:2013ava}, and NLO
electroweak~(EW) corrections for photon pair with up to two jets are also
known~\cite{Chiesa:2017gqx}. Very recently, the NNLO QCD results for $p p \to
\gamma\gamma + j$ were obtained as well~\cite{Chawdhry:2021hkp,
  Agarwal:2021vdh, Agarwal:2021grm}, and the two-loop amplitudes for $g g \to
\gamma \gamma + j$~\cite{Badger:2021imn}
were also computed.  Finally, as far
as all-order results are concerned, Sudakov logarithms of soft and collinear
origin, arising at all orders in the computation of the diphoton
transverse-momentum distribution, were resummed up to
next-to-next-to-leading-logarithmic~(NNLL) accuracy in
refs.~\cite{Balazs:2006cc, Nadolsky:2007ba, Balazs:2007hr, Cieri:2015rqa,
  Coradeschi:2017zzw}. More recently, results accurate up to N$^3$LL and
N$^3$LL$^\prime$ have also been obtained within the \Radish{}
formalism\footnote{This was shown in the plots in ref.~\cite{Alioli:2020qrd}
obtained with the \MatrixRadish{}
interface~\cite{Kallweit:2020gva}.}~\cite{Monni:2016ktx, Bizon:2017rah} and
the {\tt CuTe-MCFM} framework~\cite{Becher:2020ugp, Neumann:2021zkb},
respectively.

Nowadays, NLO Monte Carlo simulations for $pp\to\gamma\gamma$, matched to
parton showers~(PS), can be easily obtained with automated tools. A few
dedicated studies have been presented in the last decade: for instance, in
ref.~\cite{Hoeche:2009xc}, hard tree-level matrix elements with a variable
photon multiplicity were merged with a QCD+EW parton shower, allowing to take
simultaneously into account the direct and the fragmentation component. In
ref.~\cite{DErrico:2011cgc} these two components were also included in the
simulation, in the context of an exact matching of the NLO QCD corrections
with parton showers~(NLO+PS), using the \pwg{}
method~\cite{Nason:2004rx}.\footnote{NLO+PS results for diphoton production,
including the $s$-channel exchange of a Kaluza-Klein resonance, were also
obtained in ref.~\cite{Frederix:2012dp}.}

Due to the increasing precision of experimental measurements and the fact
that NNLO QCD corrections to photon-pair production are known and sizable, it
is particularly important to match parton showers with the NNLO fixed order
calculation (NNLO+PS). A few methods have been developed to obtain NNLO+PS
predictions for color-singlet production at hadron colliders: ``NNLO
reweighted'' \MiNLO$^\prime$~\cite{Hamilton:2012rf, Hamilton:2013fea}
(sometimes also denoted  NNLOPS in the literature), {\tt
  Geneva}~\cite{Alioli:2012fc, Alioli:2013hqa},
{\tt UN$^2$LOPS}~\cite{Hoche:2014uhw}, and \MiNNLOPS{}~\cite{Monni:2019whf,
  Monni:2020nks}.\footnote{Ideas for going beyond NNLO+PS accuracy have been
discussed as well (see refs.~\cite{Frederix:2015fyz, Prestel:2021vww}), and very recently
NNLO+PS matching with sector showers, for final-state parton showers, has
been outlined in ref.~\cite{Campbell:2021svd}.}  Using these methods, many
processes, that at leading order~(LO) have only two external colored legs, have been described with NNLO+PS
accuracy~\cite{Hamilton:2013fea,
  Karlberg:2014qua, Astill:2016hpa, Astill:2018ivh, Re:2018vac,
  Bizon:2019tfo, Alioli:2015toa, Alioli:2019qzz, Alioli:2020fzf,
  Alioli:2020qrd, Alioli:2021qbf, Alioli:2021egp, Cridge:2021hfr,
  Hoche:2014uhw, Hoche:2014dla, Hoche:2018gti, Monni:2019whf, Monni:2020nks,
  Lombardi:2020wju, Lombardi:2021rvg, Hu:2021rkt,
  Buonocore:2021fnj}. Recently, results for top-pair production at NNLO+PS
were also published in ref.~\cite{Mazzitelli:2020jio, Mazzitelli:2021mmm}. As
far as diphoton production is concerned, the only NNLO+PS accurate results
have been obtained with the {\tt Geneva} method in
ref.~\cite{Alioli:2020qrd}.

In this paper we study diphoton production using the \MiNNLOPS{} method,
i.e.~we build a \MiNLO$^\prime$ simulation for $pp\to \gamma\gamma$ + jet
production, and we subsequently include NNLO QCD corrections according to the
procedure proposed in refs.~\cite{Monni:2019whf, Monni:2020nks}. The
resulting generator can be used to obtain NLO+PS accurate results for $\ggj$
production, as well as to predict, with up to NNLO accuracy, observables that
are inclusive with respect to the presence of jets, such as the diphoton
invariant mass and rapidity, or the transverse momentum of the hardest and
next-to-hardest photon, retaining a consistent matching with parton showers.

Together with the main phenomenological results for diphoton
production, in the current paper we also describe a few novelties
introduced to deal with
the presence of
photons in the final state, but that could also be useful 
for other (N)NLO+PS calculations.
First of all, already for the simulation of diphoton + jet production at
NLO+PS accuracy, one needs to deal with the fact that the Born-level $\ggj$
matrix elements are divergent whenever a photon becomes soft or collinear to
a quark. In the following, we refer to these divergences as ``QED
divergences''.\footnote{Since no cuts are applied at the generation level, we
need to devise a treatment of QED-divergent regions during the generation of
the events, even though the final fiducial cuts will provide infrared-safe
results.}
We describe and implement a general way to deal with this issue in the \pwg{}
formalism.\footnote{This method has already been used in
ref.~\cite{Lombardi:2020wju}, following the suggestion of some of us.}
Furthermore, the \MiNNLOPS{} method for the process at hand requires the
evaluation of the $q\bar{q}\to\gamma\gamma$ matrix-elements up to second
order in the strong coupling constant. Such matrix elements are divergent
when the photons are collinear to the beam axis: in order to avoid
introducing any cutoff or isolation criteria at any stage of the event
generation, we have also devised a new mapping from the $pp\to \gamma\gamma +
j$ to the $pp\to \gamma\gamma$ kinematics, that preserves the direction of
one photon with respect to the beam axis,
thereby allowing for a full control of the singular regions. The combined use
of the new mapping and of the method to deal with the QED divergences allowed
us to simulate diphoton production at NNLO+PS accuracy without introducing
any generation or technical cut, as done instead in
refs.~\cite{Lombardi:2020wju, Alioli:2020qrd}.
This gives us the
advantage that, on one side, we do not have to worry about checking the
independence of the fiducial physical differential cross sections from the
generation and technical cuts, and, on the other side, we can use the same
set of generated events for any choice of fiducial cuts.

In this paper we also propose a generalization of the \MiNNLOPS{} method that
allows for greater flexibility in the choice of the renormalization and
factorization scale used in the evaluation of the non-singular term for $F$ +
1 jet production, where $F$ is the color-singlet system.  In the original
formulation of the \MiNNLOPS{} method, such term is evaluated with scales of
the order of the transverse momentum of $F$.  The prescription we introduce
here generalizes such aspect of the \MiNNLOPS{} method, allowing one to treat
this term more similarly to a fixed-order computation. Such choice turned out
to be desirable in~$\gamma\gamma$ production, for comparisons with
fixed-order results, where this term gives an important contribution to the
total cross section, unlike in previous studies, such as those in
refs.~\cite{Monni:2019whf, Monni:2020nks}, where it was small.

The paper is organized as follows: in section~\ref{sec:outline} we describe
the ingredients used to build the event generator and the details of the
novel aspects we introduce here for the first time, whose validation is
discussed in section~\ref{sec:validation}. In section~\ref{sec:results} we
show some phenomenological results, and we compare our predictions with the
ATLAS diphoton data~\cite{ATLAS:2021mbt}. Finally we give our conclusions and
outlook in section~\ref{sec:conclusions}.

\section{Outline of the calculation}
\label{sec:outline}

In this section we review the theoretical formulation of our calculation.  We
first introduce the basic notation used throughout the paper and describe the
main theoretical issues to be dealt with in the implementation of a NNLO+PS
event generator for photon pair production. Then we discuss in detail the
handling of the QED singularities and introduce the mapping used to
project the $\Phi_{\gamma\gamma j}$ kinematics onto the $\PhiB$ one.  We
conclude the section with a description of the modifications we have
introduced in the \MiNNLOPS{} formalism in order to reproduce more accurately
the results of a fixed-order calculation for distributions totally inclusive
with respect to the QCD radiation.

\subsection{Description of the process}
\label{sec:implementation}
We consider the process of direct production of a photon pair in 
proton-proton scattering
\begin{equation}
p p \,\to\, \gamma \gamma +X,
\end{equation}
with the requirement that the two photons are isolated, i.e.~each photon has
a minimum transverse momentum and is isolated with respect to the other
photon and to the final-state hadronic activity.  These requirements are
needed to make the process well-defined both from the theoretical and
the experimental point of view.
We call $\pgo$ and
$\pgt$ the momenta of the hardest and next-to-hardest photons and denote the
squared mass  and  transverse momentum
of the diphoton system as
\begin{equation}
  \label{eq:maa}
  Q^2 = m_{\gamma\gamma}^2 = \(\pgo+\pgt\)^2\!, \qquad \qquad
  \pt = \(\pgo+\pgt\)_\tr.
\end{equation}
The starting point for the  \MiNNLOPS{} method we use in this work is the NLO
differential cross section for diphoton plus one jet production
\begin{equation}
pp \,\to\, \gamma\gamma  + j\,.
\end{equation}
The matrix elements for this process at NLO in QCD have
been obtained from the automated interface of the
\pwgboxres~\cite{Jezo:2016ujg} to \OpenLoopsTwo~\cite{Buccioni:2019sur,
  Cascioli:2011va, Buccioni:2017yxi, vanHameren:2010cp,
  vanHameren:2009dr}.

Since the goal of the  \MiNNLOPS{} formalism is to reach NNLO accuracy for
inclusive observables in the diphoton system, one also needs
the two-loop amplitudes for
$q\bar{q}\to\gamma\gamma$, that have  been taken from
refs.~\cite{Anastasiou:2002zn, Catani:2013tia} and implemented into the code.

In this paper, we work in the approximation of five light quarks and neglect
the contributions given by two-loop diagrams with the massive top quark. We
consider instead the contributions given by the massive top quark in the one-loop
diagrams for the process $pp \to \gamma\gamma + j$.

\subsection{Handling of the QED divergences}
\label{sec:QEDdiv}

In this section, we describe a general way to deal with processes that
present QED divergences at the Born level within the \pwg{}
formalism.\footnote{The way the \PowhegBox{} deals with QCD divergences
at Born level has already been illustrated in several
papers,~e.g. ref.~\cite{Alioli:2010qp}.}
Before illustrating the method, we briefly review the relevant parts
of the \PowhegBox{} framework.

We start by introducing the general form of the \pwg{}~\cite{Nason:2004rx}
differential cross section for a $2 \to n$ Born process, using the notation
of refs.~\cite{Frixione:2007vw, Alioli:2010xd}. Indicating with $\mathd\Phin$
the phase space for the $n$-body final state, we write the $(n+1)$-body phase
space $\mathd\Phinpo$ in terms of $\mathd\Phin$ and three additional
radiation variables, that we collectively label $\Phirad$
\begin{equation}
\mathd\Phinpo = \mathd\Phin \, \mathd\Phirad\,.
\end{equation}
We indicate with $B(\Phin)$, $V(\Phin)$ and $R(\Phinpo)$ the Born, virtual
and real amplitudes, convoluted with the corresponding parton distribution
functions (PDFs) and multiplied by the flux factor, and we split $R$ into two
positive contributions: $R_s(\Phinpo)$, that contains all the QCD
singularities, and a QCD-finite term, $R_f(\Phinpo)$, such
that\footnote{Please notice that $R_f(\Phinpo)$ could well be set to zero,
in which case $R_s(\Phinpo) = R(\Phinpo)$, the whole real contribution.}
\begin{equation}
\label{eq:Rs+Rf}
 R(\Phinpo) = R_s(\Phinpo) + R_f(\Phinpo).
\end{equation}
In general we achieve this separation with the help of a suitable function
$F(\Phinpo)$,
such that
\begin{equation}
R_s(\Phinpo) = F(\Phinpo) \, R(\Phinpo)\,, \qquad \qquad R_f(\Phinpo)
= \lq 1-F(\Phinpo)\rq R(\Phinpo)\,. 
\end{equation}
The detailed form  for $F(\Phinpo)$ will be discussed in section~\ref{sec:F}.
When $R(\Phinpo)$ is split as in eq.~(\ref{eq:Rs+Rf}),
the \pwg{} differential cross section can be written as
\begin{eqnarray}
\label{eq:PWG_sig}
\mathd\sigma &=& \bar{B}\!\(\Phin \)\!  \lg \Delta\!\(\Phin,\ktmin\)
+ \theta\!\(\kt-\ktmin\) \Delta\!\(\Phin,\kt\) \, \frac{{
    R_s\!\(\Phinpo \)}}{B\!\(\Phin \)} \, \,\mathd \Phirad \rg
\mathd\Phin
\nonumber \\
&& {} + R_f\!\(\Phinpo \) \mathd\Phinpo\,,
\end{eqnarray}
where $\kt$ is the transverse momentum of the \pwg{} jet,
\begin{equation}
\label{eq:Bbar}
\bar{B}(\Phin) =  B(\Phin) + V(\Phin) + \int \mathd \Phirad \, { R_s(\Phin,
  \Phirad)}\,,
\end{equation}
and the \pwg{} Sudakov form factor
\begin{equation}
\label{eq:Sudakov_PWG}
 \Delta\!\(\Phin, \kt\) = \exp \lg - \int \mathd \Phirad'\,
   { \frac{{ R_s(\Phin, \Phirad')}}{B(\Phin)}} \,  \theta\!\lq
   \kt'\!\(\Phin,\Phirad'\) - \kt \rq   \rg
\end{equation}
contains only the QCD singular part of the real contributions
$R_s(\Phinpo)$. In the expressions above $\ktmin$ acts as an infrared cutoff
for unresolved radiation.

If the Born amplitude is divergent, the \PowhegBox{} applies a suppression
factor $\SB(\Phin)$ to $\bar{B}(\Phin)$ such that the product $\SB(\Phin)
\bar{B}(\Phin)$ is integrable over the entire Born phase space, and the
$\Phin$ kinematics can be generated from the distribution $\SB(\Phin)\bar{B}(\Phin)$
without the need of any hard cut. At the end, each event is given a weight
$1/\SB(\Phin)$ in order to compensate for the suppression factor.

In general, this way of regularizing divergences by means of a suppression
factor that depends on $\Phin$ can be used every time QCD and/or QED
singularities are present at the Born level.
While applying a suppression factor on $\bar{B}(\Phin)$ is enough when only
QCD singularities are present, for diphoton production we also have to deal with the QED
singularities appearing inside $R_f(\Phinpo)$.
We cannot simply suppress them through a second suppression factor
(that would be a function of $\Phinpo$) since this term is integrated
in the radiation variables inside $\bar{B}(\Phin)$, preventing one to compensate for such suppression factor, after the events have been generated.

We then proceed as follows: we take advantage of the possibility to
separate the real contributions into two terms (see
eq.~(\ref{eq:Rs+Rf})) in such a way that $R_s(\Phinpo)$ contains all the QCD
singularities but no QED ones.
Since the generation of $\Phinpo$ according to $R_f(\Phinpo)$ in
eq.~(\ref{eq:PWG_sig}) is performed with a hit-and-miss technique, we apply a
QED suppression factor $\SR(\Phinpo)$ to $R_f(\Phinpo)$, and generate
$\Phinpo$ according to the product $\SR(\Phinpo) R_f(\Phinpo)$, that is now
integrable. Finally, we multiply the weights of the generated events by the
factor $1/\SR(\Phinpo)$.\footnote{To the best of our knowledge, a similar
procedure was used for the first time in ref.~\cite{DErrico:2011cgc}.}

\subsubsection[The damping function $F$]{The damping function $\boldsymbol{F}$}
\label{sec:F}

According to the FKS method~\cite{Frixione:1995ms, Frixione:1997np}, the real
contribution $R$ is partitioned into a sum of terms $R^{\alr}$, each of them
having at most one  collinear and one soft singularity associated with one
parton

\begin{equation}
 R = \sum_{\alr} R^\alr\,.
 \end{equation}
In each $\alr$ region, the radiated parton~(r) can then become soft or be collinear to an emitting
one~(e), and we can define the damping function
\begin{equation}
F^\alr = \frac{
  \(  \frac{\displaystyle 1}{\displaystyle d_{\rm\sss  e,r} }  \)^p
  }
  {\displaystyle
 \(  \frac{\displaystyle 1}{\displaystyle d_{\rm\sss  e,r}}  \)^p
   +  \sum_{i = 1}^{n_{\rm c}} \sum_{j=1}^{n_{\gamma}}  
  \( \frac{1}{d_{c_i,\gamma_j}} \)^p }\,,
\end{equation}
where
\begin{equation}
  \label{eq:dij}
  d_{i,j} = \lg
  \begin{array}{l l}
    \displaystyle p_{\tr j}^2& \mbox{if $i$ is an initial-state
      particle,}
    \\[2mm]
    \displaystyle 2\min\!\(E_i^2,E_j^2\) \(1-\cos\theta_{ij}\) \qquad & 
    \mbox{if $i$ and $j$ are final-state particles,} 
  \end{array}
  \right.
\end{equation}
and the sum in the denominator runs over the $n_{\sss\rm c}$ massless charged
particles and the $n_{\gamma}$ photons. In eq.~(\ref{eq:dij}), $p_{\tr j}$
and $E_j$ are the transverse momentum and energy of the particle $j$, and
$\theta_{ij}$ the angle between the particles $i$ and $j$, computed in the
partonic center-of-mass frame.\footnote{In our simulation, we have set
$p=2$.}

Every contribution $R^\alr$ to the real amplitude is then split into
two terms
\begin{equation}
R^{\alr} = R^{\alr}_{s} + R^{\alr}_{f}\,,
\end{equation}
where 
\begin{equation}
R^{\alr}_{s} \equiv F^{\alr} R^{\alr}, \qquad \qquad\qquad R^{\alr}_{f}
\equiv \( 1- F^{\alr} \) R^\alr\,.
\end{equation}
In the limit where the radiated parton is soft or collinear to the emitter
$d_{\rm\sss e,r}$ is small, and $F^{\alr}$ tends to 1, so that all the QCD
singularities are in $R^{\alr}_{\sss\rm s}$.  When instead the photon
$\gamma_j$ becomes soft or collinear to a massless charged particle $c_i$,
the term $d_{c_i,\gamma_j}$ is small, and $F^{\alr}$ tends to 0, so that
$R^{\alr}_{\sss\rm s}$ is free from QED singularities.

\subsubsection[The suppression factors $\SB$ and
  $\SR$]{The suppression factors $\boldsymbol{\SB}$ and
  $\boldsymbol{\SR}$}
\label{sec:S_B-S_R}
We have chosen as  suppression factor $\SB(\Phin)$ introduced in
section~\ref{sec:QEDdiv} the following expression
\begin{eqnarray}
  \label{eq:supp_fact_Born}
\SB &=& \frac{\( p^2_{\tr j}\)^a}{ \( p^2_{\tr j}\)^a + \(\bar{p}^2_{\tr
    j}\)^a}
\,\frac{\( p^2_{\tr \gamma_1}\)^a}{ \( p^2_{\tr \gamma_1}\)^a + \(\bar{p}^2_{\tr \gamma}\)^a}
\,\frac{\( p^2_{\tr \gamma_2}\)^a}{ \( p^2_{\tr \gamma_2}\)^a +
  \(\bar{p}^2_{\tr \gamma}\)^a}
\nonumber\\[2mm]
&& {} \times
\frac{\( R^2_{j \gamma_1}\)^a}{ \( R^2_{j \gamma_1}\)^a + \(\bar{R}^2_{j \gamma}\)^a}
\,\frac{\( R^2_{j \gamma_2}\)^a}{ \( R^2_{j \gamma_2}\)^a + \(\bar{R}^2_{j \gamma}\)^a},
\end{eqnarray}
where $p_{\tr i}$ is the transverse momentum of the particle $i$ with
respect to the beam axis, $R_{ij}$ the angular distance between
the particles $i$ and $j$ in the azimuth-pseudorapidity plane
\begin{equation}
    R_{ij} =\sqrt{\(\eta_i-\eta_j\)^2+\(\phi_i-\phi_j\)^2},
\end{equation}
and the barred quantities are arbitrary constant parameters needed to define
the transition point (from zero to one) of each factor in the product.  The
first three term of eq.~(\ref{eq:supp_fact_Born}) suppress the limit where
the final-state parton or the two photons are soft or collinear to the beam
axis, while the last two terms suppress the regions where the photons are
collinear to the final-state parton. The power $a$ does not need to be the
same for all the terms, but we have chosen the common value $a = 1$ for
simplicity. The expression for $\SR(\Phinpo)$ can be obtained by generalizing
eq.~(\ref{eq:supp_fact_Born}), for the case with an extra jet.

When the \MiNNLOPS{} formalism is used, the QCD initial-state singularity is
already regularized by the Sudakov form factor from the resummation
formalism, and the first term in eq.~(\ref{eq:supp_fact_Born}) is no longer
needed. We give more details in section~\ref{sec:S_B-S_R_MiNNLO}.

\subsection[The $\PhiBj \to \PhiB$ mapping]{The $\boldsymbol{\PhiBj} \to \boldsymbol{\PhiB}$ mapping}
\label{sec:mapping}

When applying the \MiNNLOPS{} formalism, we need a mapping from the
$\gamma\gamma j$ phase space to the $\gamma\gamma$ one. In fact, as it will
be recalled in the next section, the Sudakov form factor (eq.~(\ref{eq:Sud}))
and the $D$ term (eq.~(\ref{eq:Dterms})), at the core of the \MiNNLOPS{}
method, are functions of $\PhiB$, as they contain the
$q\bar{q}\to\gamma\gamma$ amplitudes. One needs then to evaluate these quantities
while integrating over the $\Phi_{\gamma\gamma j}$ phase space upon which the
\pwg{} $\bar{B}$ function depends, thereby requiring a smooth 
$\Phi_{\gamma\gamma j}\to \PhiB$ mapping.

The $q\bar{q}\to\gamma\gamma$ amplitudes are singular when the photons are
collinear to the beam axis. From the physics point of view, one does not
expect to evaluate such amplitudes arbitrarily close to their singularities,
as such phase space points are discarded by the request of the presence of
two isolated photons in the final state.  Since in the \MiNNLOPS{} formalism
the $\PhiB$ kinematics is obtained by a mapping from the $\Phi_{\gamma\gamma
  j}$ one, we need to make sure that we never get too close to the singular
regions, and without having to introduce an explicit cut on the transverse momentum of
the single photons in the $\PhiB$ kinematics.

We have built such a mapping without having to introduce any cutoff or
isolation criteria, at any stage of the event generation, as done in other
Monte Carlo simulations dealing with photons~\cite{Alioli:2020qrd,
  Lombardi:2020wju}. 
The mapping that we propose is such that the direction of one of the photons
with respect to the beam axis in the laboratory frame, for a given phase space
point $\Phi_{\gamma\gamma j}$, is preserved in the projected point $\PhiB$. As
a consequence, a point in $\PhiB$ with small $\pt$ always comes from a
projection of a point in $\Phi_{\gamma\gamma j}$ where at least one photon is
close to the beam axis, and this configuration is suppressed by the factor
$\SB$ of eq.~(\ref{eq:supp_fact_Born}).
A detailed derivation of such a mapping is given in
appendix~\ref{app:projection}.

\subsection{\MiNNLOPS{} differential cross section}
\label{sec:MiNNLO}
In this section, after briefly recalling the general method, we
discuss the modifications that we have introduced in the definition of
the \MiNNLOPS{} differential cross section given in
refs.~\cite{Monni:2019whf, Monni:2020nks}.
Following the conventions introduced in those papers,
we write the $\pt$ spectrum of the cross section
for diphoton production as
\begin{equation}
\label{eq:sig_s_f}
  \frac{\mathd\sigma}{\mathd\PhiB\mathd \pt} =
  \frac{\mathd\sigmas}{\mathd\PhiB\mathd \pt} +
  \frac{\mathd\sigmaf}{\mathd\PhiB\mathd \pt},
\end{equation}
where $\sigmas$ (also called singular contribution in the following)
is obtained from the resummation of logarithmically enhanced
contributions at small-$\pt$, while $\sigmaf$ (also called non-singular term) is the
difference between the fixed-order differential cross section
and the truncated
perturbative $\as$ expansion of $\sigmas$, till the second order.
The singular contribution can be written as
\begin{equation}
  \label{eq:sig_s}
  \frac{\mathd\sigmas}{\mathd\PhiB\mathd \pt} = \frac{\mathd}{\mathd
    \pt} \lg \exp\!\lq -\tilde{S}\!\(\PhiB,\pt\) \rq \! \Lum\!\(\PhiB,
  \pt\)\rg,
\end{equation}
where the Sudakov form factor is given by
\begin{equation}
  \label{eq:Sud}
  \tilde{S}\!\(\PhiB, \pt\) = 2\int_{\pt}^{Q}\frac{\mathd q}{q}
  \lg  A\Big(\! \as\!\(q\)\!\Big) \ln\frac{Q^2}{q^2} +
  \tilde{B}\Big(\! \as\!\(q\)\!\Big) \rg
\end{equation}
and $\Lum$ is a function of the PDFs, of the $q \bar{q} \to \gamma
\gamma$ Born, one- and two-loop matrix elements, and of the collinear
coefficient functions (see ref.~\cite{Catani:2012qa}). The functions
$A$ and $\tilde{B}$ in eq.~(\ref{eq:Sud}) can be written as
\begin{eqnarray}
  A(\as) &=& \left(\abar\right) A^{(1)} + \left(\abar\right)^2
  A^{(2)}+ \left(\abar\right)^3 A^{(3)}\,,
  \\
  \tilde{B}(\as) &=& \left(\abar\right) B^{(1)} + \left(\abar\right)^2
  \tilde{B}^{(2)}\,,
\end{eqnarray}
with
\begin{equation}
  \tilde{B}^{(2)} = B^{(2)} + \beta_0 H^{(1)} + 2\zeta_3\(A^{(1)}\)^2,
\end{equation}
where $H^{(1)}$ is the ratio between the one-loop and the Born $q \bar{q} \to
\gamma \gamma$ amplitudes in the \MSB{} subtraction scheme, and introduces a
dependence on the phase-space kinematics in $\tilde{B}^{(2)}$. The
coefficients $A^{(1)}$, $A^{(2)}$, $A^{(3)}$, $B^{(1)}$, and $B^{(2)}$ for
quark-initiated processes are collected, for example, in
ref.~\cite{Monni:2019whf}. More details can be found in
appendix~\ref{app:scale_dep}.

The non-singular contribution $\mathd\sigmaf$ is instead given by the
difference
\begin{equation}
  \label{eq:sig_f}
  \frac{\mathd\sigmaf}{\mathd\PhiB\mathd \pt} =
\left.
\frac{\mathd\sigma^{\rm \sss NLO}_{\gamma\gamma j}}{\mathd\PhiB\mathd \pt}
\right|_Q
- \abarmu{Q} \!
\left[\frac{\mathd\sigmas}{\mathd\PhiB\mathd \pt}\right]^{(1)}_Q
- \left(\abarmu{Q}\right)^{\!2}\!
\left[\frac{\mathd\sigmas}{\mathd\PhiB\mathd \pt}\right]^{(2)}_Q,
\end{equation}
where $\mathd\sigma^{\rm \sss NLO}_{\gamma\gamma j}$ is the NLO
differential cross section for $\gamma\gamma j$ production, and $\lq
\mathd\sigmas \rq^{(i)}$ is the $i$-th order of the expansion of
$\mathd\sigmas$ in the strong coupling constant. In this paper, we use
the notation $[X]^{(i)}$ for the coefficient of the $i$-th term in the
perturbative expansion of the quantity $X$. The above difference is
integrable, since the expansion of $\sigma_s$ cancels the
non-integrable terms of $\mathd\sigma^{\rm \sss NLO}_{\gamma\gamma j}$
in the $\pt \to 0$ limit.
At variance with refs.~\cite{Monni:2019whf, Monni:2020nks,
  Lombardi:2020wju, Lombardi:2021rvg, Mazzitelli:2020jio}, we have
chosen to set the renormalization and factorization scales $\mur$ and
$\muf$ in $\mathd\sigmaf$ to $Q$, instead of $\pt$. This is why we
have added the subscript $Q$ to the quantities appearing in
eq.~(\ref{eq:sig_f}). While formally any scale choice would be
legitimate for evaluating this term, since the difference would be of
order ${\cal O}(\as^3)$, beyond the accuracy of our calculations,
setting the scale to $Q$ allows to follow more closely what is
typically adopted in fixed-order calculations, thereby allowing for a
more accurate comparison with the latter.
Moreover, the choice of the central scale also plays a role in the
estimation of the theoretical uncertainties. In this case, since the
size of the non-singular contribution is numerically relevant, setting
the scale to $\pt$ would result in an unnatural enhancement of the
scale-variation bands.
We will give more quantitative details in
section~\ref{sec:DY_and_Higgs}, where we compare the new $\mu = Q$ and the
$\mu = \pt$ scale choices for $\mathd\sigmaf$ for the previously
discussed cases of Drell-Yan and Higgs boson production.

Following refs.~\cite{Monni:2019whf, Monni:2020nks}, the singular
contribution of eq.~(\ref{eq:sig_s}) can be rewritten as
\begin{equation}
\label{eq:expand_diff}
  \frac{\mathd\sigmas}{\mathd\PhiB\mathd \pt} = \exp\!\lq
  -\tilde{S}\!\(\PhiB, \pt\) \rq \! D\!\(\PhiB, \pt\),
\end{equation}
where
\begin{equation}
\label{eq:Dterms}
D\!\(\PhiB, \pt\)  \equiv -\frac{\mathd \tilde{S}\!\(\PhiB,\pt\)}{\mathd \pt} \, \Lum\!\(\PhiB, \pt\)
+ \frac{\mathd \Lum\!\(\PhiB, \pt\)}{\mathd \pt}\,.
\end{equation}
By expanding eq.~(\ref{eq:expand_diff}) in $\as(Q)$, we can write
\begin{eqnarray}
\label{eq:dsigs1}
\left[\frac{\mathd\sigmas}{\mathd\PhiB\mathd \pt}\right]^{(1)}_Q &=&
\Big[ D\!\(\PhiB, \pt\) \Big]^{(1)}_Q
\\
\label{eq:dsigs2}
\left[\frac{\mathd\sigmas}{\mathd\PhiB\mathd \pt}\right]^{(2)}_Q &=& 
\Big[ D\!\(\PhiB, \pt\) \Big]^{(2)}_Q - \lq \tilde{S}\!\(\PhiB,\pt\) \rq^{(1)}
\Big[ D\!\(\PhiB, \pt\) \Big]^{(1)}_Q
\end{eqnarray}
where all the terms are evaluated at $\mur=\muf=Q$.

The $D\!\(\PhiB, \pt\)$ term in eq.~(\ref{eq:expand_diff}) has a
formal expansion 
\begin{eqnarray}
\label{eq:D_expans_pt}
D\!\(\PhiB, \pt\) &=&
\frac{\as(\pt)}{2\pi} \lq D\!\(\PhiB, \pt\) \rq^{(1)} +
\(\frac{\as(\pt)}{2\pi}\)^2 \lq D\!\(\PhiB, \pt\) \rq^{(2)}
\nonumber\\
&& \mbox{} + 
\(\frac{\as(\pt)}{2\pi}\)^3 \lq D\!\(\PhiB, \pt\) \rq^{(3)} + \ldots
\end{eqnarray}
and, at difference with eqs.~(\ref{eq:dsigs1}) and~(\ref{eq:dsigs2}), all
the terms are evaluated at $\mur=\muf=\pt$.\footnote{We stress the fact that
we do not compute separately the $\lq D\!\(\PhiB, \pt\) \rq^{(i)}$ terms, but
we compute numerically the whole function $D$, as discussed, for the
first time, in ref.~\cite{Monni:2020nks}.}
The explicit expression of these terms are collected in
appendix~\ref{app:scale_dep}, together with the expressions of the other
ingredients needed in the \MiNNLOPS{} formulae.

We would like to stress that the cancellation of the divergences associated
with the small $\pt$ limit in eq.~(\ref{eq:sig_f}) is numerically
challenging.
For this reason, and following the \MiNLO{} original approach,  we have chosen to
multiply $\mathd\sigmas$ by a Sudakov form factor, after adding an
additional term to preserve the nominal $\as^2$ accuracy
\begin{equation}
\label{eq:sigf_times_sud}
\frac{\mathd\sigmaf}{\mathd\PhiB\mathd \pt} \, \to \, \exp\!\lq
-{\bar{S}}\!\(\pt\) \rq \lg
\frac{\mathd\sigmaf}{\mathd\PhiB\mathd \pt} +
\(\frac{\as(Q)}{2\pi}\)^2 \lq {\bar{S}}\!\(\pt\) \rq^{(1)} \lq
\frac{\mathd\sigmaf}{\mathd\PhiB\mathd \pt}\rq^{(1)} \rg.
\end{equation}
At difference with~\cite{Monni:2019whf, Monni:2020nks,
  Lombardi:2020wju, Lombardi:2021rvg, Mazzitelli:2020jio}, where
$\bar{S}=\tilde{S}$, we apply a Sudakov
form factor with only the two leading terms, i.e.
\begin{equation}
  \bar{S}\!\(\pt\) = 2 \int_{\pt}^{Q} \frac{dq}{q}
  \frac{\as\!\(q\)}{2\pi} \lq  A_1\log\!\(\frac{Q^2}{q^2}\) +
    B_1 \rq.
\end{equation}
We stress that the two approaches are equivalent up to $\ord{\as^2}$.

Summarizing, the \MiNNLOPS{} differential
cross section for $pp\to \gamma\gamma + X$ production can be written
as
\begin{eqnarray}
\label{eq:dsig_final}
\frac{\mathd\sigma_{\sss \gamma\gamma j}}{\mathd\PhiBj}
&=&
\exp\!\lq -\tilde{S}\!\(\PhiB,\pt\) \rq \! D\!\(\PhiB, \pt\)  \Fcorr
\nonumber\\
&+&
\exp\!\lq -{\bar{S}}\!\(\pt\) \rq
\!\Bigg\{\!
\left.
\frac{\mathd\sigma^{\rm \sss NLO}_{\gamma\gamma j}}{\mathd\PhiBj}
\right|_Q
+
 \(\frac{\as(Q)}{2\pi}\)
 \lq {\bar{S}}\!\(\pt\) \rq^{(1)}
 \left.
 \frac{\mathd\sigma^{\rm \sss  LO}_{\gamma\gamma j}}{\mathd\PhiBj}
\right|_Q
\nonumber\\
&&
\mbox{}\hspace{15mm}
- \lq \abarmu{Q} \!
\Big[ D\!\(\PhiB, \pt\)\! \Big]^{(1)}_Q
+ \left(\abarmu{Q}\right)^{\!2}\!
\Big[ D\!\(\PhiB, \pt\)\! \Big]^{(2)}_Q \rq \!\Fcorr
\!\!\Bigg\},
\nonumber\\
\end{eqnarray}
where we have introduced the symbol ${\mathd\sigma^{\rm \sss LO}_{\gamma\gamma
    j}}/{\mathd\PhiBj}$ to denote the LO differential cross section for $p p \to
\gamma \gamma j$ production,
and the function $\Fcorr$, needed to spread the contributions
proportional to the $D(\PhiB, \pt)$ terms over the entire $\PhiBj$ phase
space. It has the property that, given an arbitrary function
$G\!\(\PhiB,\pt\)$,
\begin{equation}
\int \mathd\PhiBj \, G\!\(\PhiB,\pt\) \Fcorr = \int
\mathd\PhiB \, \mathd\pt \, G\!\(\PhiB,\pt\).
\end{equation}
For the explicit expression we have used for $\Fcorr$ we refer to
ref.~\cite{Monni:2019whf}.

Equation~(\ref{eq:dsig_final}) is the main result of this section and it
summarizes the novel aspects we introduced to the \MiNNLOPS{} method,
as can be evinced by comparing it against,~e.g., eq.~(3.7) of
ref.~\cite{Monni:2019whf}. In the rest of the manuscript, we will
denote the \MiNNLOPS{} results obtained according to
eq.~(\ref{eq:dsig_final}) with the acronym \FOatQ. We also
recall that the mapping introduced in section~\ref{sec:mapping}
guarantees that all the $\PhiB$-dependent terms in
eq.~(\ref{eq:dsig_final}) are
evaluated in kinematic points away from the singular regions.

By rewriting the NLO differential cross section for $\gamma\gamma j$
production in the following compact form
\begin{equation}
  \mathd\sigma^{\rm \sss NLO}_{\gamma\gamma j} = \(B + V \) {\mathd\PhiBj} +
  \(R_{\sss s} + R_{\sss f}\){\mathd\PhiBjj}\,,
\end{equation}
and introducing the  \MiNNLOPS{}
improved $\bar{B}$ function
\begin{eqnarray}
  \bar{B}\!\(\PhiBj\)_{\rm\sss MiNNLO_{PS}}\! &=&\!
 \exp\!\lq -{\tilde{S}}\!\left(\PhiB, \pt\right)\rq
 D\!\(\PhiB, \pt\) \Fcorr
\nonumber \\
&& {}\! + \exp\!\lq -{\bar{S}}\!\left(\pt\right)\rq \Bigg\{\!\!
\lq 1 + \frac{\as(Q)}{2\pi}  \lq {\bar{S}}\!\left(\pt\right) \rq^{(1)} 
\rq  B \big|_Q     +  V\big|_Q  +
\int \mathd\Phi_\rad \,  R_s\big|_Q 
\nonumber \\
&& {}\! - \left[\! \left(\frac{\as(Q)}{2\pi}\right) \!  \Big[ D\!\(\PhiB,
    \pt\)\! \Big]^{(1)}_Q + \left(\abarmu{Q}\right)^{\!2}\!  \Big[
    D\!\(\PhiB, \pt\)\! \Big]^{(2)}_Q \right] \!\Fcorr\!\! \Bigg\},
\nonumber\\
\end{eqnarray}
our final expression for  the differential cross section $\mathd\sigma_{\sss
  \rm \gamma\gamma}$ reads
\begin{eqnarray}
  \label{eq:MiNNLOPS+Powheg_formula}
  \mathd\sigma_{\sss \rm \gamma\gamma} &=&  \bar{B}\!\(\PhiBj\)_{\rm\sss MiNNLO_{PS}}
  \Bigg\{ \Delta\!\left(\PhiBj, \ktmin\right)
\mathd\PhiBj  +  \theta\!\left(\kt- \ktmin\right)
\Delta\!\left(\PhiBj,\kt\right)
\left. \frac{R_{\sss s}}{B}\right|_\kt \mathd\PhiBjj \Bigg\}
\nonumber \\
&& {} + R_{\sss f} \big|_Q \, \mathd\PhiBjj\,.
\end{eqnarray}

\subsubsection{The suppression factors with \MiNNLOPS{}}
\label{sec:S_B-S_R_MiNNLO}
As already discussed in section~\ref{sec:S_B-S_R}, we do not need to suppress
the small $\pt$ region of the first jet while the \pwgboxres{} integrates the
$\bar{B}\!\(\PhiBj\)_{\rm\sss MiNNLO_{PS}}$ function over the whole phase
space, since the presence of the \MiNNLOPS{} Sudakov form factor suppresses
such region.
In this case, the Born suppression
factor $\SB$ in eq.~(\ref{eq:supp_fact_Born}) can be replaced by
\begin{equation}
  \label{eq:supp_fact_Born_MiNNLO}
S_{\sss \rm B}^{\rm\sss MiNNLO} = \frac{\( p^2_{\tr \gamma_1}\)^a}{ \( p^2_{\tr \gamma_1}\)^a + \(\bar{p}^2_{\tr \gamma}\)^a}
\,\frac{\( p^2_{\tr \gamma_2}\)^a}{ \( p^2_{\tr \gamma_2}\)^a +
  \(\bar{p}^2_{\tr \gamma}\)^a}
\frac{\( R^2_{j \gamma_1}\)^a}{ \( R^2_{j \gamma_1}\)^a + \(\bar{R}^2_{j \gamma}\)^a}
\,\frac{\( R^2_{j \gamma_2}\)^a}{ \( R^2_{j \gamma_2}\)^a + \(\bar{R}^2_{j \gamma}\)^a}\,.
\end{equation}

\subsubsection{Scale settings in the small-$\pt$ limit and modified logarithms}
\label{sec:scales}

We freeze the renormalization and factorization scales at values below $Q_0 =
2$~GeV to avoid the Landau pole and further non-perturbative effects,
connected with the PDF evolution to lower scales.
We stress that this scale does not act as a cutoff in the integration over
the physical space
but 
only enters in the evaluation of the singular
contribution in eq.~(\ref{eq:expand_diff}), since all the other terms
in our formulation of the \MiNNLOPS{} differential cross section are evaluated
at the hard scale $Q$.
We also highlight that, due to the presence of an overall Sudakov form
factor, the dependence of the differential cross 
section on $Q_0$ is strongly suppressed.\footnote{
We have found no visible dependence on $Q_0$ of our results by lowering its value
down to values of 1~GeV.}

In addition, following what was done in ref.~\cite{Monni:2019whf, Monni:2020nks}, we
smoothly turn off the contribution of the logarithms in the $D$ functions at
large transverse momentum with the replacement
\begin{equation}
  \log\frac{Q}{\pt} \, \to \, \frac{1}{p} \log\!\( 1 + \(\frac{Q}{\pt}\)^p\),
  \end{equation}
so that the $\pt \to 0$ limit remains unaffected, while at $\pt > Q$, the
modified logarithm tends to zero. In our simulation we have set
$p=6$.\footnote{We have explicitly checked that the phenomenological results
we present in section~\ref{sec:results} are independent of the exact value of $p$. In fact, we
have changed this value in the range from 1 to 9 without obtaining any appreciable
difference in the final results.}

\section{Validation}
\label{sec:validation}
In this section we discuss the validation of our \MiNNLOPS{} predictions.  We
first briefly present the setting of the physical and technical parameters of
the calculation and the isolation criterion used to define the diphoton
process.  We then study the effects that the modifications to the \MiNNLOPS{}
differential cross section described in section~\ref{sec:MiNNLO} have on two
previous implementations of the method (i.e.~the Drell-Yan and Higgs
production processes). Finally, we present a validation of the \MiNNLOPS{}
results for diphoton production against the fixed-order distributions
produced with the public version of the \Matrix{} code.

\subsection{Physical and technical parameters and photon isolation criterion}
\label{sec:physical_parameters}

The phenomenological results presented in this paper have been
obtained for a proton-proton collider with a hadronic center-of-mass
energy of 13~TeV. We have used the \LHAPDF~\cite{Buckley:2014ana}
parton distribution function set \verb|NNPDF31_nnlo_as_0118| and the
evolution of $\as$ provided by the same package. The electromagnetic
coupling for the final-state photons has been set to $\alpha =
{1}/{137}$, and the mass of the top quark to $m_{\sss\rm top} =
173.2~{\rm GeV}$.

The fixed-order results have been obtained with the public
  version of the
\Matrix{} code~\cite{Grazzini:2017mhc, Catani:2011qz,
  Cascioli:2011va,Denner:2016kdg, Anastasiou:2002zn, Catani:2012qa,
  Catani:2007vq}, setting the central renormalization and
factorization scales equal to the invariant mass of the diphoton
system $\mur = \muf = Q$. The uncertainty band has been estimated
via a seven-point scale variation obtained by
multiplying and dividing the central renormalization and factorization scales by a
factor of 2.

We apply the photon
isolation prescription of ref.~\cite{Frixione:1998jh} to the two
final-state photons. For each photon, we compute the angular distance
$R_{i\gamma}$ with respect to the $i$-th final-state parton. We
discard the event unless, for every photon and every $R<R_c$,
\begin{equation}
  \label{eq:fx_isolation}
  \sum_{i=1}^{n_{\rm \sss i}} \pti  \, \theta \!\(R-R_{i\gamma}\) < \ETmax \,  \chi(R),
\end{equation}
where $n_{\rm\sss i}$ is the number of final-state partons, $\pti$ is
the transverse momentum of $i$ with respect to the beam, and
\begin{equation}
  \chi(R) = \( \frac{1 - \cos R}{1 - \cos \Rcone}\)^n\!.
\end{equation}
In our analysis, we have set
\begin{equation}
  \ETmax = 4~{\rm GeV}, \qquad\qquad \Rcone = 0.4\,, \qquad\qquad  n=1\,.
\end{equation}
In addition, the two photons have to fulfill
\begin{equation}
  \label{eq:kinematic_cuts}
  \pTgamone > \pTgamonemin\,,
  \qquad\qquad
  \pTgamtwo > \pTgamtwomin\,,
  \qquad\qquad
  \mgamgam > \mgamgammin\,,
\end{equation}
where $\pTgamone$ and $\pTgamtwo$ are the transverse momenta of the
hardest and next-to-hardest photons,  $\mgamgam$ is the diphoton
mass, and
\begin{equation}
  \label{eq:pheno_cuts}
  \pTgamonemin = 25~{\rm GeV},
  \qquad\qquad
  \pTgamtwomin = 22~{\rm GeV},
  \qquad\qquad
  \mgamgammin = 25~{\rm GeV}.
\end{equation}
The values of the barred quantities in eq.~(\ref{eq:supp_fact_Born_MiNNLO})
and of the power $a$ have been set to
\begin{equation}
\bar{p}_{\tr \gamma} = 22~{\rm GeV},\qquad\qquad  \bar{R}_{j \gamma} = 0.4\,,
\qquad\qquad a = 1\,.
\end{equation}
In addition, the \pwgboxres{} code has been run with the flag {\tt doublefsr}
set to 1. This parameter was first introduced in ref.~\cite{Nason:2013uba},
and we refer the interested reader to this paper for further details.

\subsection{Validation for Drell-Yan and Higgs boson production}
\label{sec:DY_and_Higgs}

In this section, we compare the results obtained with the 
original \MiNNLOPS{} method
used
in ref.~\cite{Monni:2020nks} for Drell-Yan and Higgs boson production against
those obtained with the new formulation spelled in
section~\ref{sec:MiNNLO}. As discussed in that section, 
one expects the new formulation to yield results compatible with the original
method for processes where the size of non-singular corrections is small compared
to the total cross section.

An estimate of the size of the non-singular correction can be obtained by comparing
the total NNLO cross section against the integral of the first term on the
right-hand side of eq.~(\ref{eq:dsig_final}) over the full phase space.  The
results obtained for diphoton, Drell-Yan and Higgs boson production are the
following
\begin{equation}
  \begin{array}{ll}
    \sigma_{\sss\rm NNLO}^{\gamma\gamma} = 155.7 \pm 1.0~{\rm pb}\,,\qquad \qquad  &
    \sigmas^{\gamma\gamma}=55.7 \pm 0.6~{\rm pb}\,,
    \\[2mm] 
    \sigma_{\sss\rm NNLO}^{\rm\sss DY}=1919\pm 1~{\rm pb}\,,  &
    \sigmas^{\rm \sss DY} = 1904\pm 3~{\rm pb}\,,
    \\[2mm]
    \sigma_{\sss\rm NNLO}^{\rm\sss H}= 39.64 \pm 0.01~{\rm pb}\,,  &
    \sigmas^{\rm \sss H} = 34.03 \pm 0.07~{\rm pb}\,.
  \end{array}
\end{equation}
For diphoton production, $\sigmas^{\gamma\gamma}$ contributes to only about
one third of the total cross section, at difference with Drell-Yan and Higgs
boson production, thereby justifying the choices made in this paper.

\begin{figure}[htb!]
  \begin{center}
    \includegraphics[width=0.49\textwidth]{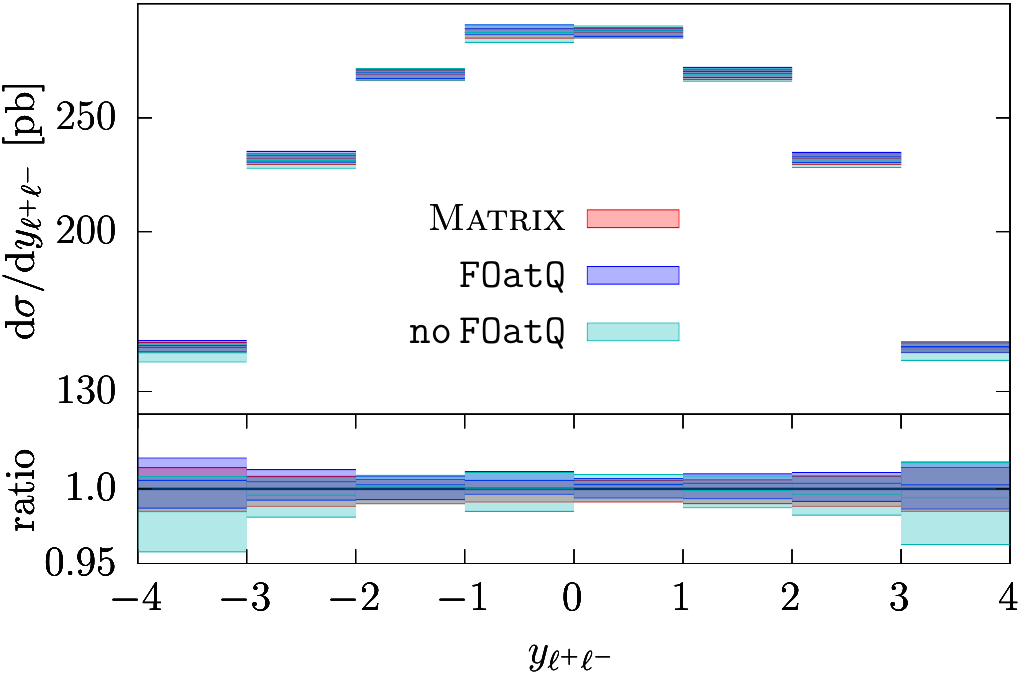}\ \
    \includegraphics[width=0.49\textwidth]{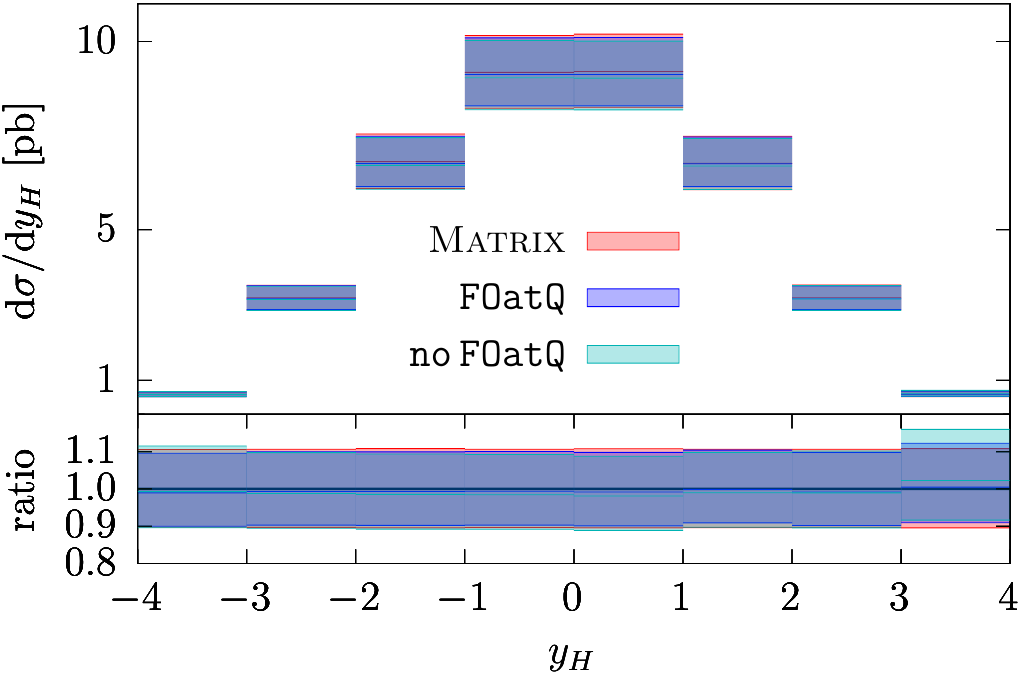}
  \end{center}
  \caption{Comparison between fixed-order predictions from \Matrix{}
    against the two \MiNNLOPS{} results (at ``Les Houches'' level)
    obtained using the original formulation of the method (labelled as
    \noFOatQ) and the new one presented in this
    work (labelled as \FOatQ). In the left pane we show the
    rapidity distribution of the dilepton system in Drell-Yan production,
    whereas in the right pane we show the Higgs boson rapidity.
  The ratios with respect to the \Matrix{} results are also shown in the lower
  panel. }
  \label{fig:validation}
\end{figure}
As a further validation, in figure~\ref{fig:validation} we compare the
rapidity distribution of the color singlet for Drell-Yan and Higgs boson
production. We show the distributions obtained with the original formulation,
where the finite terms are evaluated at $\mu=\pt$ (labelled as \noFOatQ), and
the new one, where such terms are evaluated at $\mu=Q$, with $Q$ the
invariant mass of the color singlet (labelled as \FOatQ).  The \MiNNLOPS{}
results shown in the figure are obtained after generating the \Powheg{}
hardest emission, i.e.~at the ``Les Houches'' level. In the plots we also
show the NNLO results from \Matrix{}, obtained setting $\mu=Q$. For this
comparison only, we use the same PDF sets as those used in
ref.~\cite{Monni:2020nks}.
In the lower inserts of the figure we plot the ratio of the displayed
distributions with respect to the \Matrix{} result.

The curves show a very good agreement between the NNLO and the \MiNNLOPS{}
results obtained with either formulations, both for the central scale and the
uncertainty band. In particular, since the Drell-Yan process features a very
small perturbative uncertainty band, the remarkable agreement between the
NNLO and \FOatQ{} curves displayed in the left pane of
figure~\ref{fig:validation} is a robust validation of the new formulation of
the \MiNNLOPS{} method.

\subsection{Diphoton production: comparison with \Matrix}
\label{sec:comparison_with_matrix}

In this section we validate the differential cross section of eq.~(\ref{eq:dsig_final}) 
against the fixed-order NNLO one implemented
within the public version of the \Matrix{} code.
We expect the two results to agree up to terms beyond the NNLO accuracy.
The \Matrix{} results have
been obtained setting the slicing parameter $r_{\sss\rm cut} =
0.0005$ and for a central scale choice $\mur=\muf=Q$. Using the isolation
criteria and the fiducial cuts reported in
section~\ref{sec:physical_parameters}, the total cross section computed by
\Matrix{} and \MiNNLOPS{} are in agreement within the statistical errors, and
they read respectively
\begin{equation}
  \sigma_{\rm\sss tot}^{\rm \sss Matrix} = 155.7 \pm 1.0~{\rm pb}\,,
  \qquad\qquad
  \sigma_{\rm\sss tot}^{\rm \sss MiNNLO} = 154.9 \pm 0.2~{\rm pb}\,.
\end{equation}

\begin{figure}[htb!]
  \begin{center}
    \includegraphics[width=0.49\textwidth]{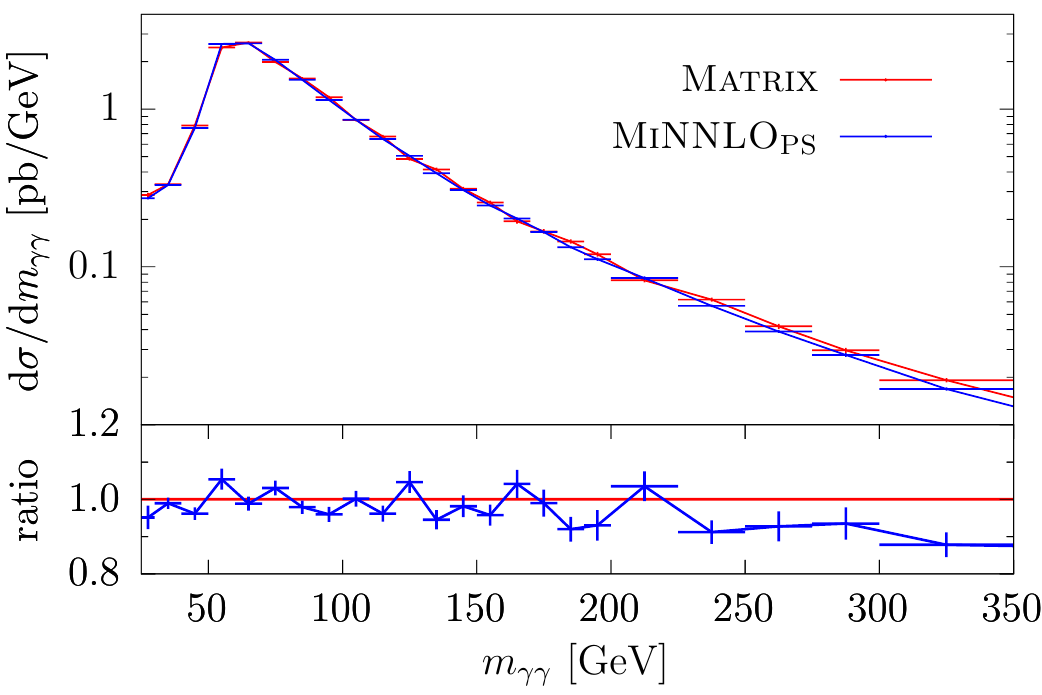}\ \
    \includegraphics[width=0.49\textwidth]{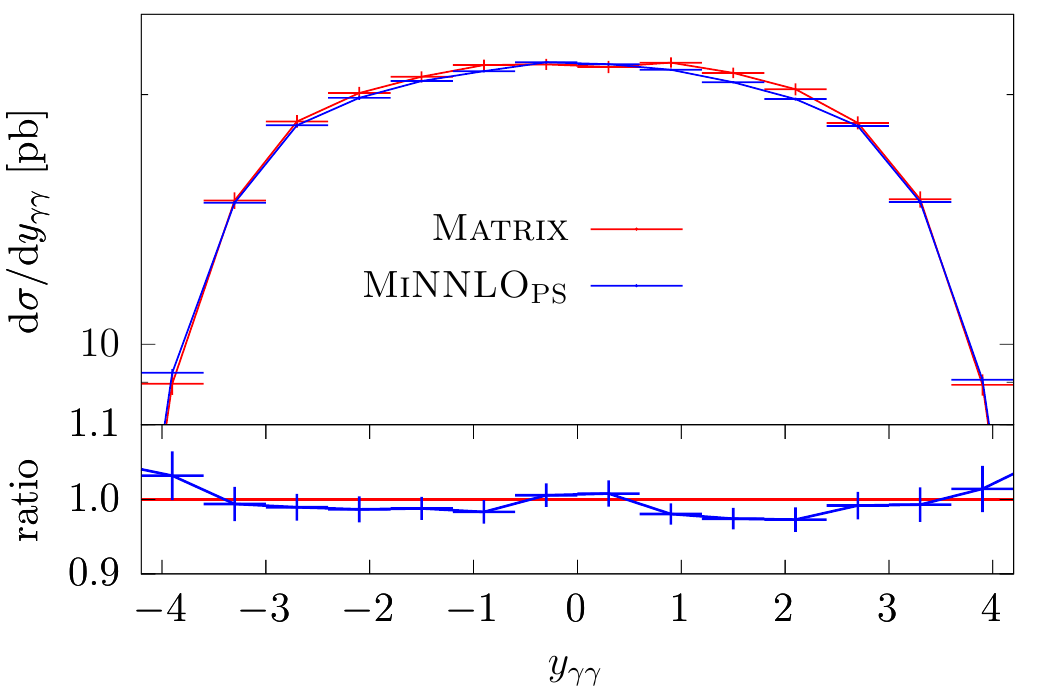}
  \end{center}
  \caption{Comparison between fixed-order results from \Matrix{} and
    \MiNNLOPS{} for the diphoton invariant mass and rapidity, for the default
     central value of the renormalization and factorization scales. The
    statistical errors are also shown.}
  \label{fig:MiNNLO_NNLO_aa}
\end{figure}
In figure~\ref{fig:MiNNLO_NNLO_aa} we plot the distributions for the
invariant mass and rapidity of the diphoton system computed with \Matrix{}
and \MiNNLOPS{}, in the top panes, and the ratio of the two curves in the
lower ones. The figure shows a reasonably good agreement between the two
curves within the statistical errors.  We ascribe the mild difference in the
high invariant-mass region to effects beyond the nominal accuracy of our
result, which can differ from the purely fixed-order \Matrix{} one by
higher-order effects, present in eq.~(\ref{eq:dsig_final}).
We have explicitly verified that, by applying a transverse momentum cut on
the diphoton system, the same trend is also present when we compare the exact
fixed-order NLO result for $\aaj$ production against the \MiNNLOPS{} one:
when we move away from the region of small transverse momentum of the
diphoton system that is affected by large logarithms, we still observe a mild
difference between the central values of the two distributions, but still within
the scale-variation bands.

\begin{figure}[htb!]
  \begin{center}
    \includegraphics[width=0.49\textwidth]{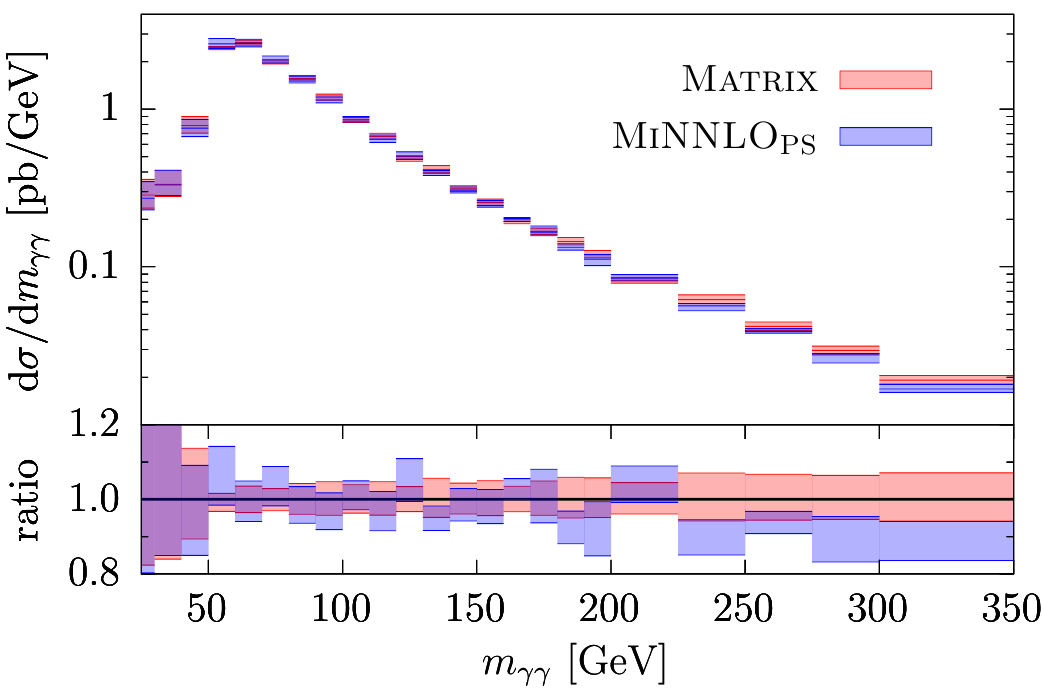}\ \
    \includegraphics[width=0.49\textwidth]{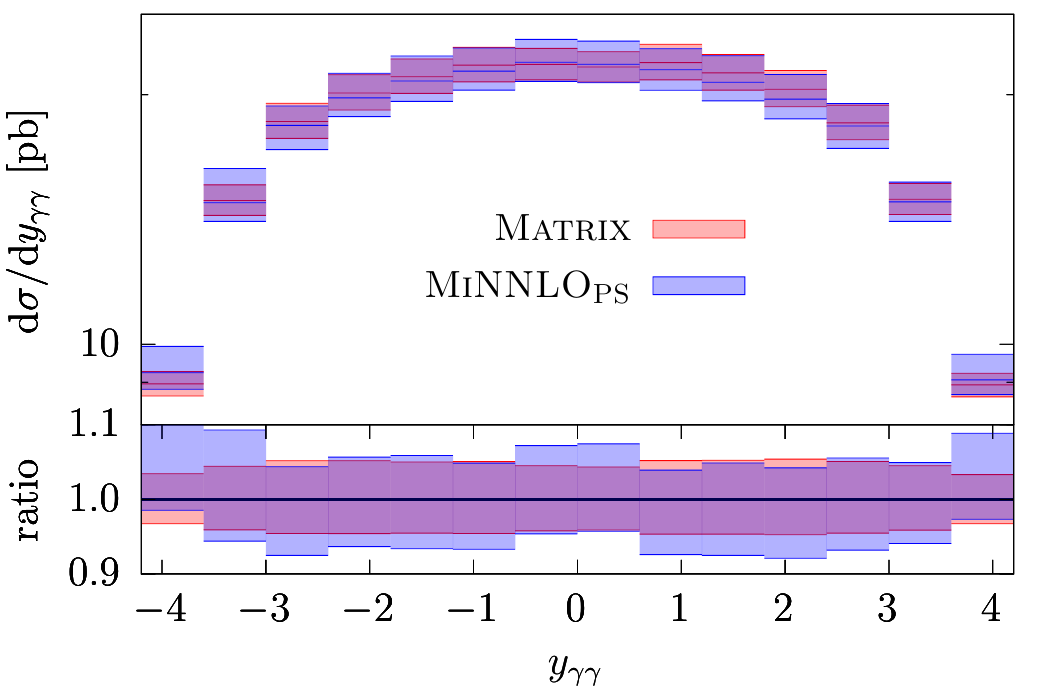}
  \end{center}
  \caption{Comparison between fixed-order results from \Matrix{} and
    \MiNNLOPS{} for the diphoton invariant mass and rapidity, as in
    figure~\ref{fig:MiNNLO_NNLO_aa}. Uncertainty bands from scale
    variations are also shown.}
  \label{fig:MiNNLO_NNLO_aa_band}
\end{figure}

\begin{figure}[htb!]
  \begin{center}  
    \includegraphics[width=0.49\textwidth]{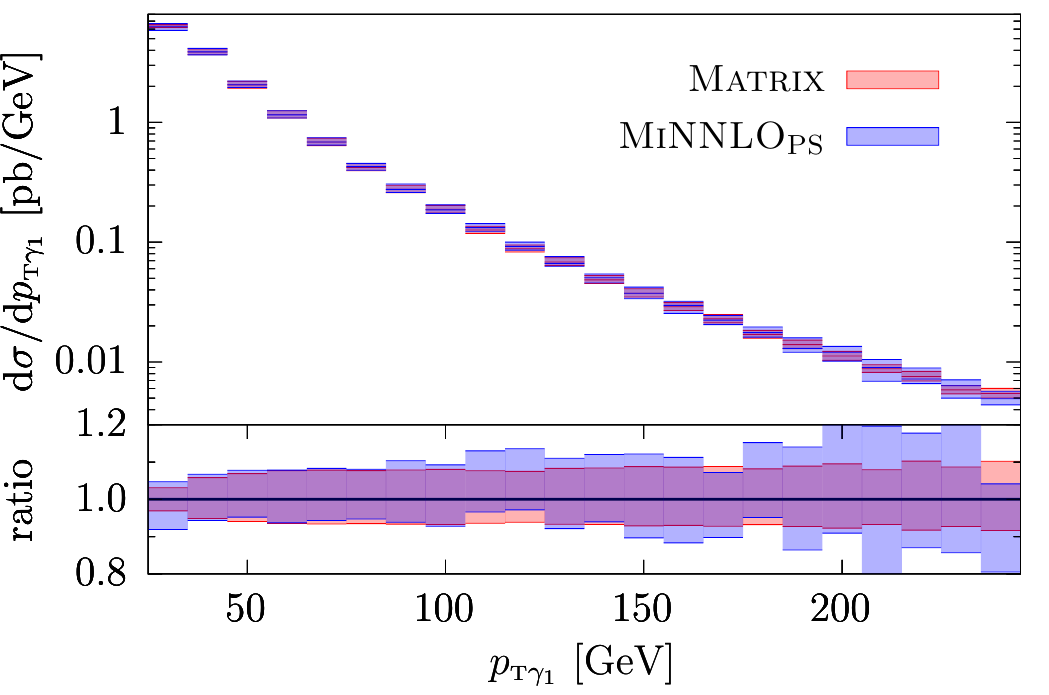}\ \
    \includegraphics[width=0.49\textwidth]{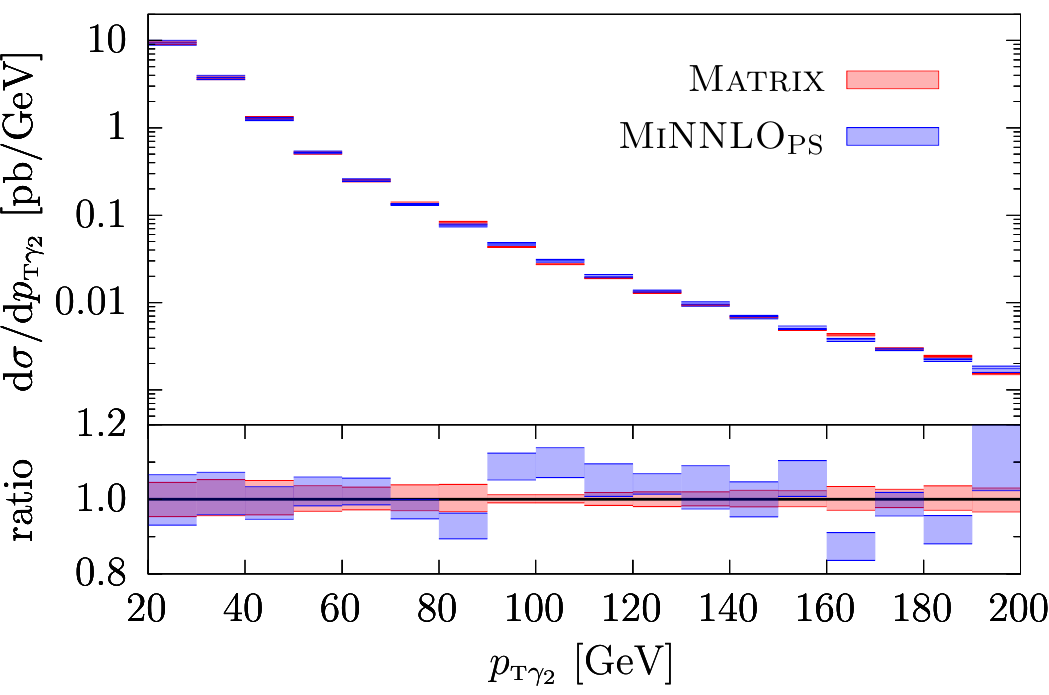}
  \end{center}
  \caption{Comparison between fixed-order results from \Matrix{} and
    \MiNNLOPS{} for the transverse momentum of the hardest and
    next-to-hardest photon. Uncertainty bands from scale
    variations are also shown.}
  \label{fig:MiNNLO_NNLO_pta_band}
\end{figure}

The theoretical uncertainties on the \Matrix{} and \MiNNLOPS{} total cross
sections, estimated through a seven-point scale variation obtained by
multiplying and dividing the central renormalization and factorization scales by a
factor of 2, are given by\footnote{\Matrix{} provides also extrapolated
values for the total cross section down to $r_{\sss\rm cut} = 0$, that, for
the central scale, is equal to $\sigma_{\rm\sss tot}^{\rm \Matrix} = 153.9
\pm 1.9$~pb.  We notice that, with the extrapolation process, the associated
statistical error is larger than in the non-extrapolated one. For comparison
with eq.~(\ref{eq:scal_var_XS}), the extrapolated scale variation band
is $\pm 4$\%.}
\begin{equation}
  \label{eq:scal_var_XS}
  \sigma_{\rm\sss tot}^{\rm \sss \Matrix} = 155.7\, ^{+5\%}
  _{-4\%}~{\rm pb}
  \qquad\qquad
  \sigma_{\rm\sss tot}^{\rm \sss MiNNLO} = 154.9 \, ^{+6\%}
  _{-5\%}~{\rm pb}\,,
\end{equation}
and are in good agreement.
In figures~\ref{fig:MiNNLO_NNLO_aa_band} and~\ref{fig:MiNNLO_NNLO_pta_band}
we compare the scale variation bands obtained from the two codes for the
invariant mass and rapidity distributions of the diphoton system and the
transverse momentum of the hardest and next-to-hardest photon.
We find an overall good agreement among the different curves, and compatible
size for the scale-variation bands.

\section{Phenomenological results}
\label{sec:results}

After validating the implementation of the \MiNNLOPS{} differential cross
section for diphoton production given in eq.~(\ref{eq:dsig_final}), in this
section we present some distributions of physical interest obtained from the
generated events, before and after passing them through a parton shower.  We
have generated about 16 million events without any generation cut, except for
a minimum invariant mass of the diphoton system of 10~GeV.\footnote{We point
out that such generation cut has no effect on the final kinematic
distributions if the fiducial cut on the diphoton invariant mass at the
analysis stage is greater than it.}
As stressed previously, except for this last constraint on the invariant
mass, no other cuts have been imposed, so that these events can be used to
make predictions with arbitrary fiducial cuts.

\subsection{Results at partonic level}
\label{sec:partonic}
In this section we compare the \Matrix{} results against those obtained after
generating the \pwg{} hardest emission of
eq.~(\ref{eq:MiNNLOPS+Powheg_formula}), often denoted as ``Les Houches
events''~(LHE), and labelled in the figures as ``\MiNNLOPS~(LHE)''.
\begin{figure}[htb!]
  \begin{center}
    \includegraphics[width=0.49\textwidth]{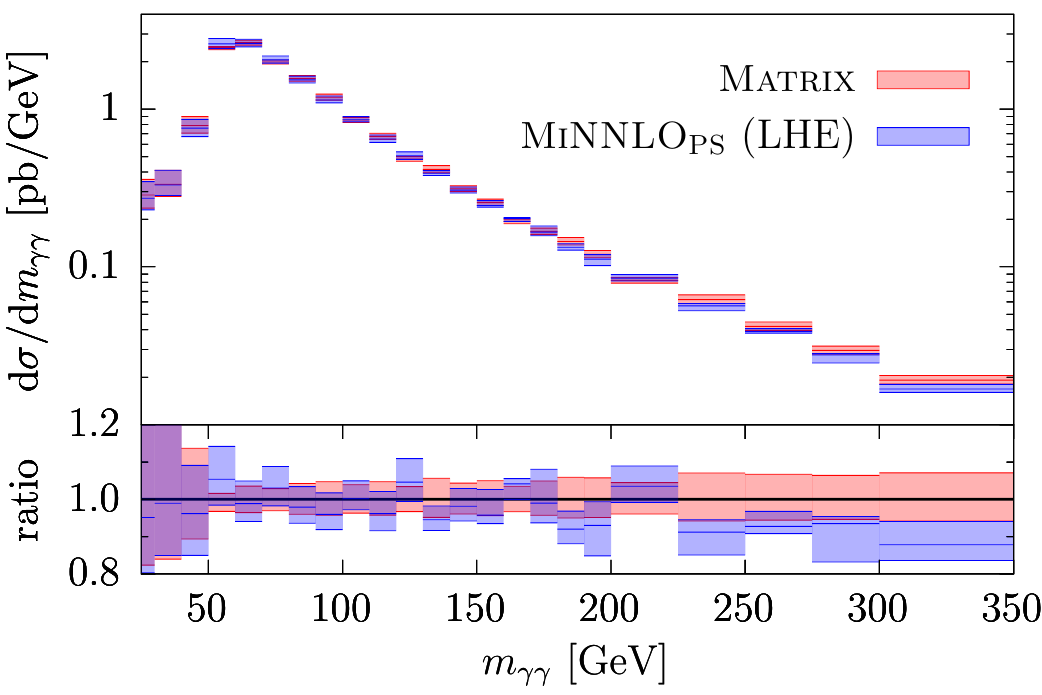}\ \
    \includegraphics[width=0.49\textwidth]{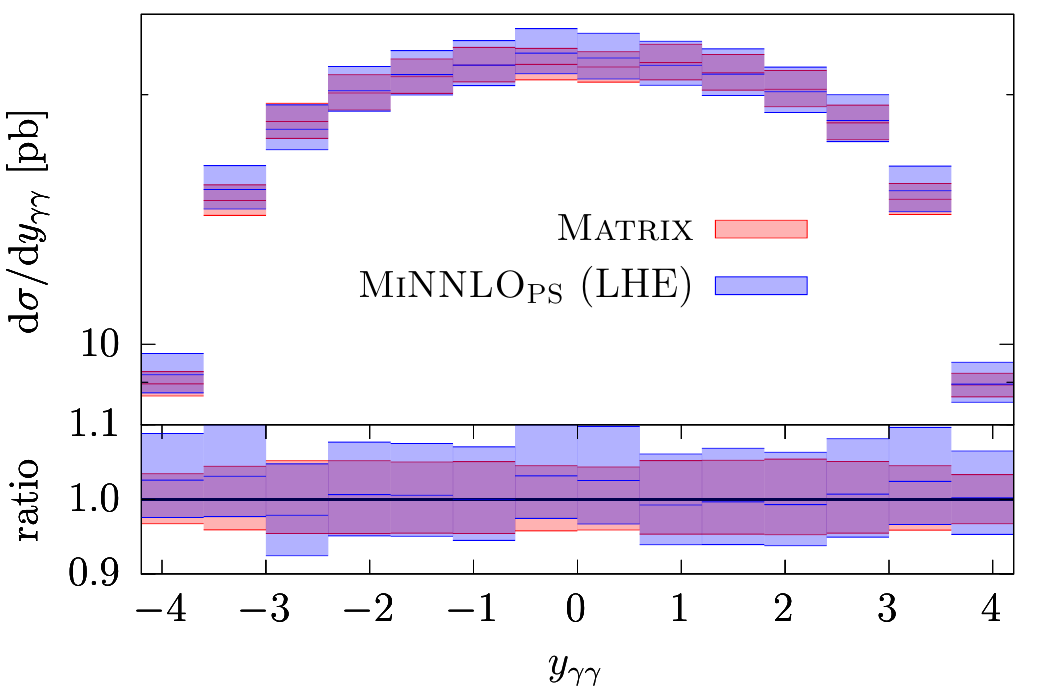}
  \end{center}
  \caption{Comparison between the distributions obtained from the
    fixed-order calculation from \Matrix{} and the results after the
    generation of the hardest emission, for the diphoton invariant
    mass and rapidity.}
  \label{fig:NNLO_LH_aa}
\end{figure}

\begin{figure}[htb!]
  \begin{center}  
    \includegraphics[width=0.49\textwidth]{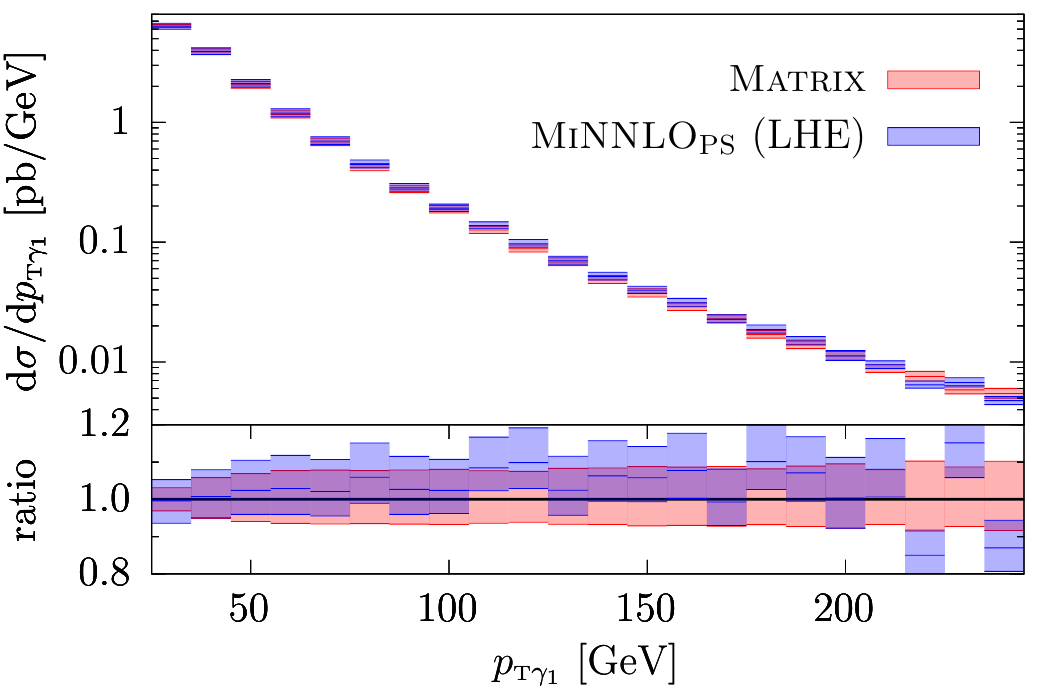}\ \
    \includegraphics[width=0.49\textwidth]{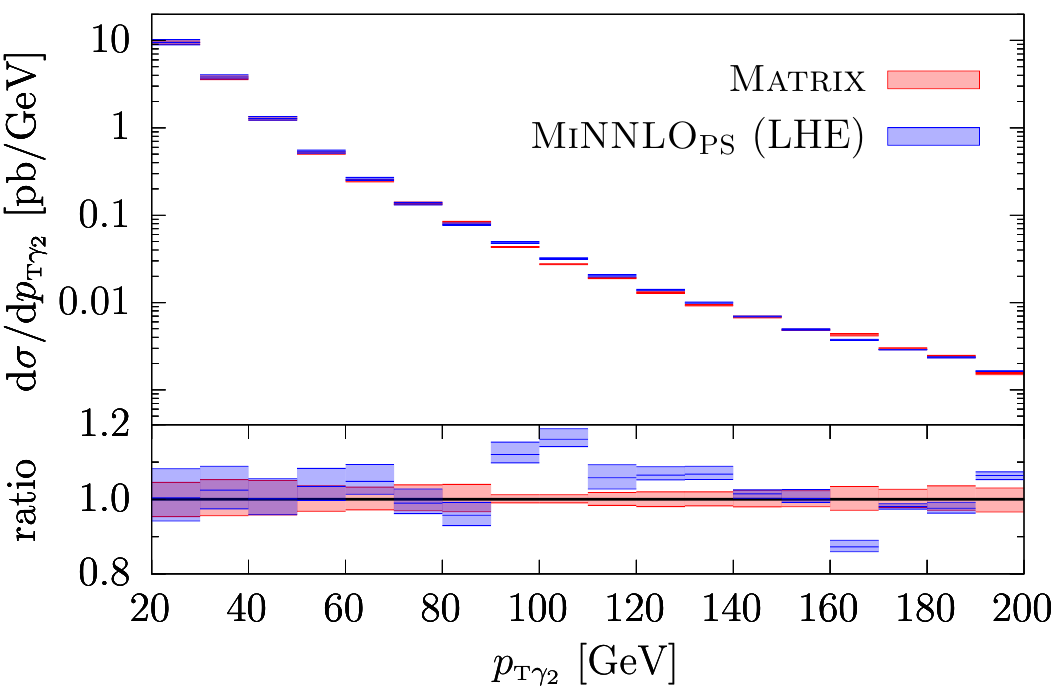}
  \end{center}
  \caption{Comparison between the distributions obtained from the
    fixed-order calculation from \Matrix{} and the results after the
    generation of the hardest emission, for the transverse momentum of
    the hardest and next-to-hardest photon.}
  \label{fig:NNLO_LH_pta}
\end{figure}
In figure~\ref{fig:NNLO_LH_aa} we show results for the invariant mass and
the rapidity of the diphoton system, whereas in
figure~\ref{fig:NNLO_LH_pta} we display the transverse momentum of the
hardest and next-to-hardest photon.
We find  good agreement between the \Matrix{} and the
\MiNNLOPS{}~(LHE) predictions, both for the central value and for the
uncertainty bands due to scale variations. Notice that, at variance
with similar comparisons for other processes not involving photons
(like Drell-Yan or massive diboson production), this is not trivial
due to the presence of an isolation criterion  in the
definition of the process.

\subsection{Results after parton showering}
\label{sec:shower}
In this section we show results obtained after the completion of the parton
shower performed by \PythiaEight{}~\cite{Sjostrand:2014zea,
  Sjostrand:2006za}. We rely on the \PythiaEight{} interface to the
\pwgboxres{}, provided by the {\tt main31} configuration file, distributed
with \PythiaEight{}.  The results presented in this section are obtained
after switching off multi-parton interactions, QED radiation and
hadronization effects, using the Monash tune~\cite{Skands:2014pea}.
We set the \PythiaEight{} parameter {\tt POWHEG:pThard} to 2~(i.e.~we use the
prescription introduced in section~4 of ref.~\cite{Nason:2013uba}), and the
{\tt SpaceShower:dipoleRecoil} to 1.

\begin{figure}[htb!]
  \begin{center}
    \includegraphics[width=0.49\textwidth]{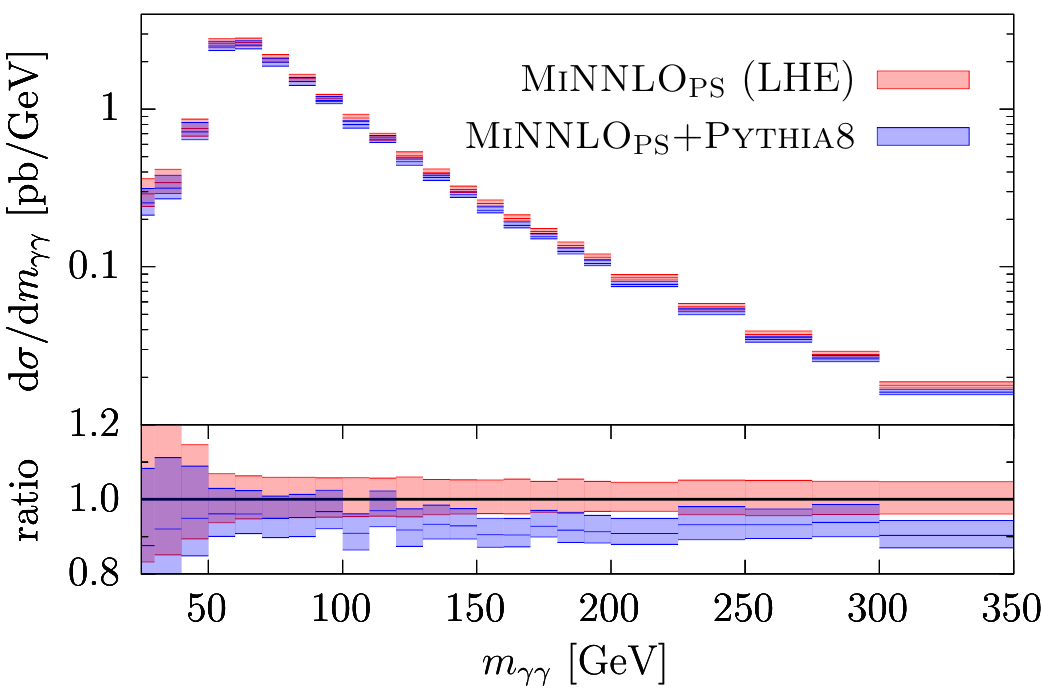}\ \
    \includegraphics[width=0.49\textwidth]{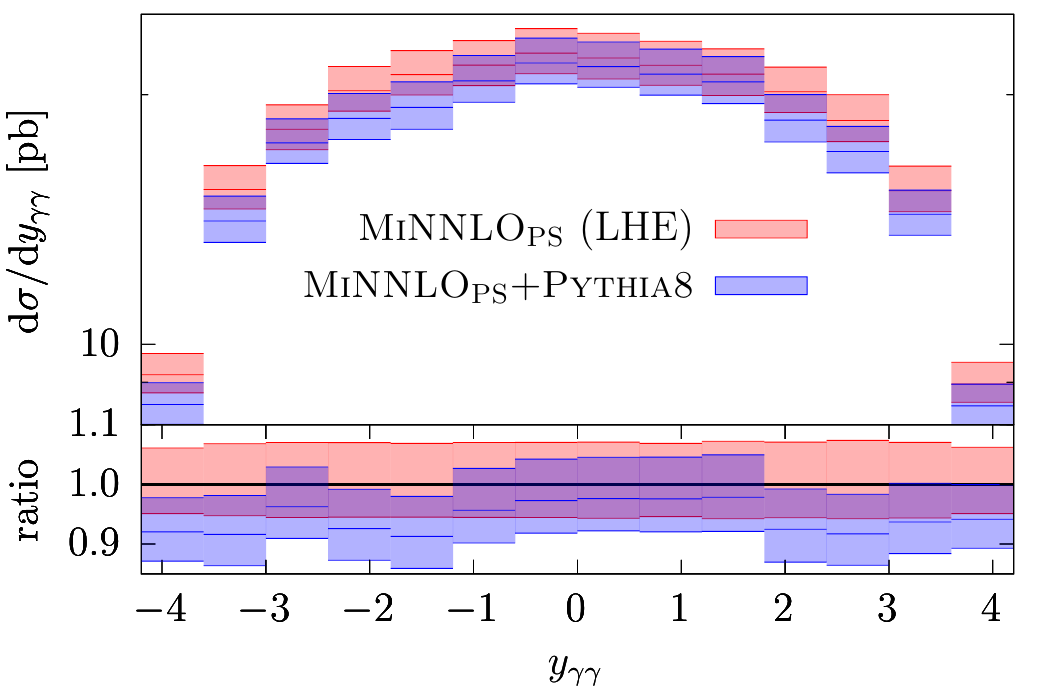}
  \end{center}
  \caption{Comparison between the distributions obtained from the
    \pwg{} events before and after the \PythiaEight{} parton shower
    for the diphoton virtuality and rapidity.}
  \label{fig:LH_PY8_aa}
\end{figure}

\begin{figure}[htb!]
  \begin{center}  
    \includegraphics[width=0.49\textwidth]{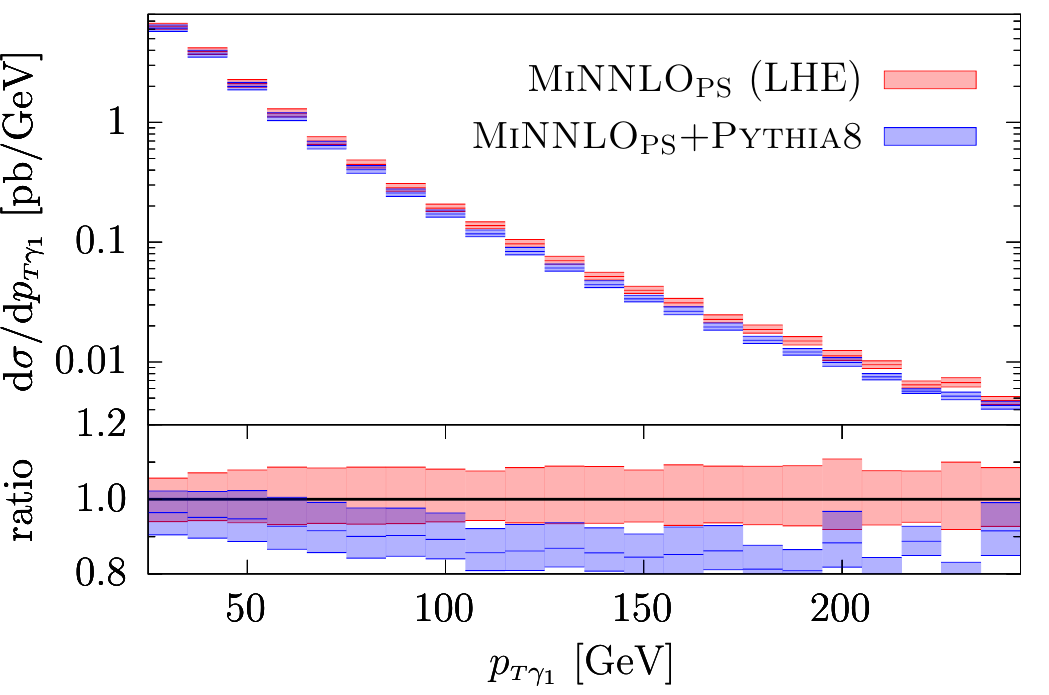}
    \includegraphics[width=0.49\textwidth]{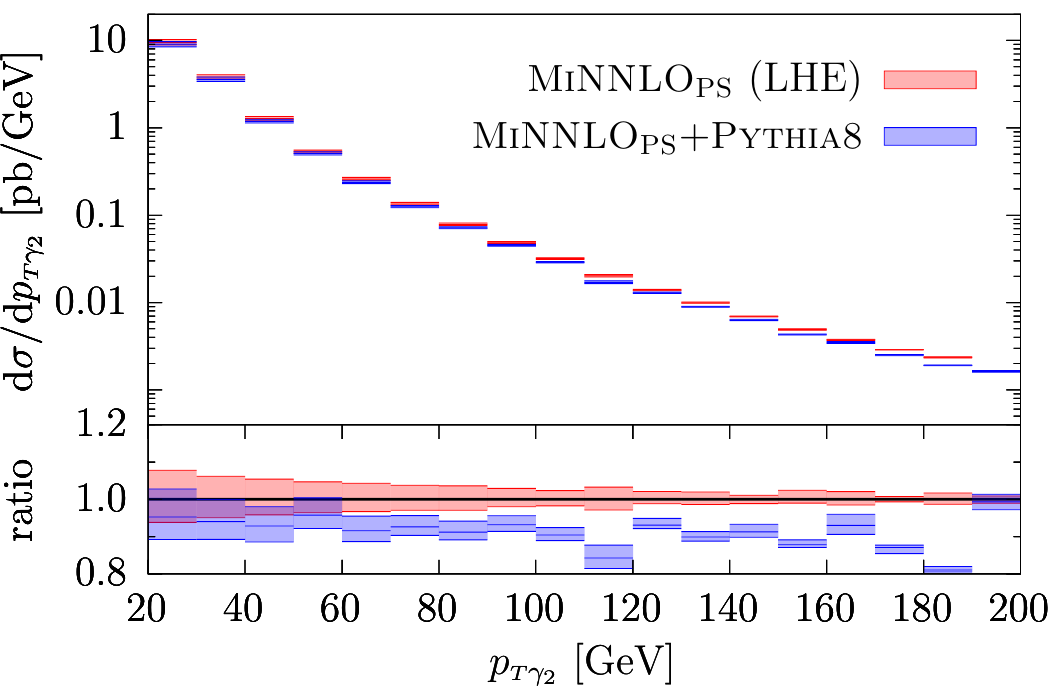}
  \end{center}
  \caption{Comparison between the distributions obtained from the
    \pwg{} events before and after the \PythiaEight{} parton shower
    for the transverse momentum of the hardest and next-to-hardest
    photon.}
  \label{fig:LH_PY8_pta}
\end{figure}
In figures~\ref{fig:LH_PY8_aa} and~\ref{fig:LH_PY8_pta}, we present the same
distributions of the previous sections, but after the \PythiaEight{} shower.
We observe a reduction of around 5--10\% of the differential cross sections,
compared to the \MiNNLOPS~(LHE), with the largest values in the region of
high transverse momenta of the two photons.  We ascribe this behaviour to the
fact that, with the increased multiplicity of the partonic/hadronic activity
due to the \PythiaEight{} shower, the photons in the showered events are less
likely to satisfy the isolation criteria, giving a smaller cross section.  We
observed such pattern also for similar distributions computed with a
completely independent code for diphoton production, built with the standard
\Powheg{} NLO+PS formalism, reassuring us that the above interpretation is
correct, and that this behaviour is not due to some \MiNNLOPS{} features.
We leave further investigation on this aspect for future studies.

\subsection{Comparison with ATLAS results}
\label{sec:atlas}

\def\yypT{\ensuremath{p_{\perp,\gamma\gamma}}\xspace}
\def\yypT{\ensuremath{{\pt}_{\gamma\gamma}}\xspace}
\def\yyphiStar{\ensuremath{\phi_{\eta}^*}\xspace}
\def\yypiMDphi{\ensuremath{\pi - \Delta\phi_{\gamma\gamma}}\xspace}
\def\yypTt{\ensuremath{a_{\perp\gamma\gamma}}\xspace}

In this section we compare our results with those presented by the ATLAS
Collaboration in ref.~\cite{ATLAS:2021mbt} obtained at 13~TeV.
In order to make a fair comparison with the experimental data, we have
added to the \MiNNLOPS{} diphoton contribution, presented in this
paper, the leading-order contribution coming from the gluon-induced
production of a photon pair through a closed quark loop (denoted by
$gg\to\gamma\gamma$ in the following), which is of the same order of
the NNLO corrections and is further enhanced by the particularly
sizable gluon luminosity.  As we consider this contribution only at
leading order, we simulate it at LO+PS accuracy, setting the
renormalization and factorization scales, and the upper limit for the
transverse momentum of the subsequent shower evolution, equal to the
invariant mass of the diphoton system.  The analytic amplitudes for
this process have been taken from ref.~\cite{Bern:2002jx} and
implemented in the \RES{} framework, neglecting the top-quark loop
contribution~\cite{Dicus:1987fk}.
This contribution amounts for at most a few percent in certain regions of the
kinematic distributions we are plotting. Its inclusion would improve the agreement
with data.

The analysis of our events is performed using
\rivet~\cite{Bierlich:2019rhm}, with the same \PythiaEight{} settings
as those in the previous section. The fiducial volume is defined by
the following requirements
\begin{equation*}
  \pTgamone > 40~{\rm GeV},
  \qquad
  \pTgamtwo > 30~{\rm GeV},
  \qquad
  \Delta R_{\gamma\gamma} > 0.4\,,
\end{equation*}
\begin{equation}
  \label{eq:atlas_cuts}
  |y_\gamma| < 1.37\,, \qquad 1.52 < |y_\gamma| < 2.37.
\end{equation}
We do not report here the exact photon-isolation criteria which are
described in detail in section~4.1 of ref.~\cite{ATLAS:2021mbt}.

\begin{figure}[htb!]
  \begin{center}  
    \includegraphics[width=0.49\textwidth]{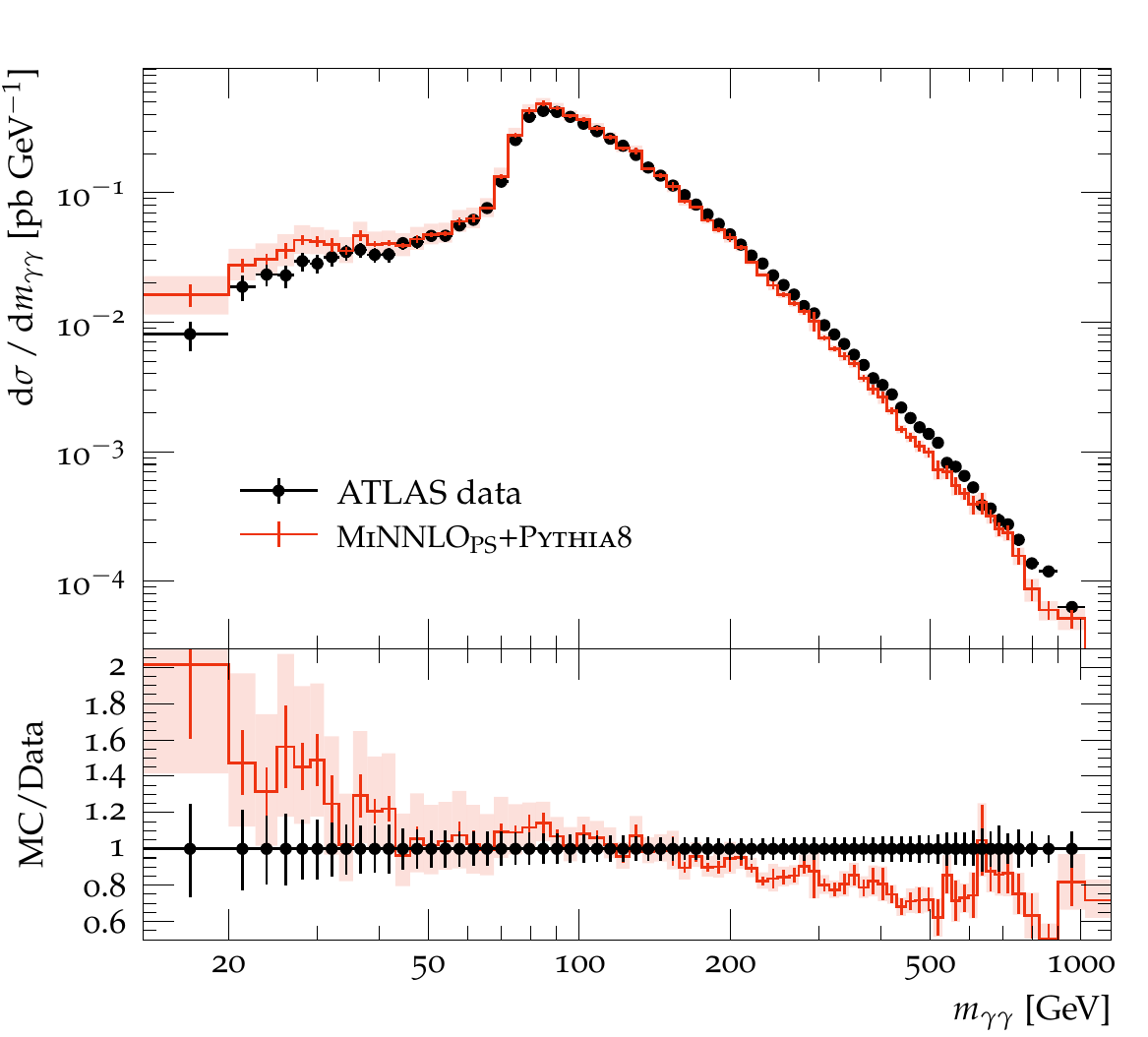}
    \includegraphics[width=0.49\textwidth]{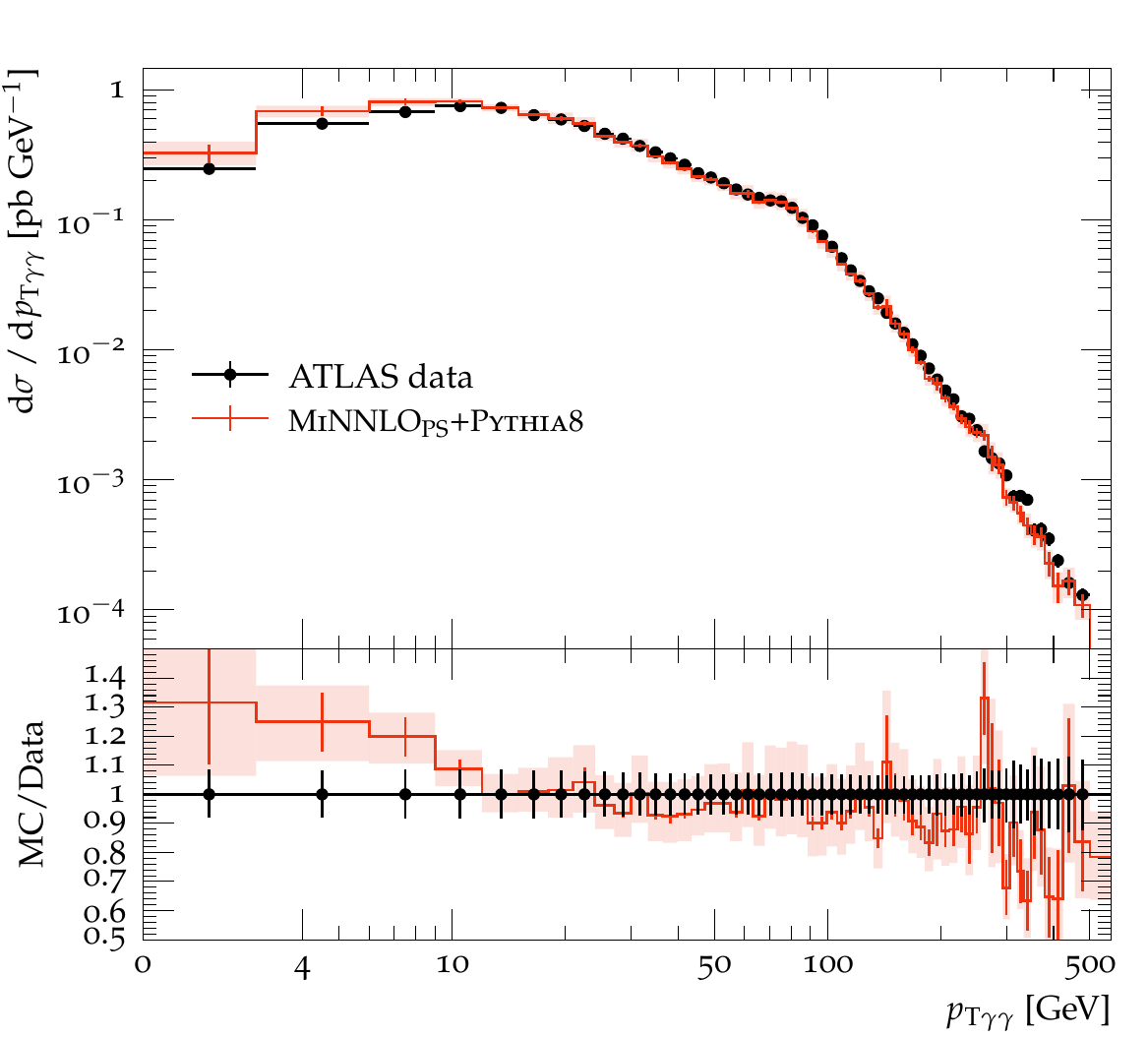}
  \end{center}
  \caption{Comparison between the ATLAS data and the distributions obtained
    with \MiNNLOPS{}+\PythiaEight{}, for the diphoton invariant mass and its
    transverse momentum. Scale-variation bands are also shown, together with
    the statistical errors of the central value.}
  \label{fig:ATLAS_yy_m_pt}
\end{figure}

\begin{figure}[htb!]
  \begin{center}  
    \includegraphics[width=0.49\textwidth]{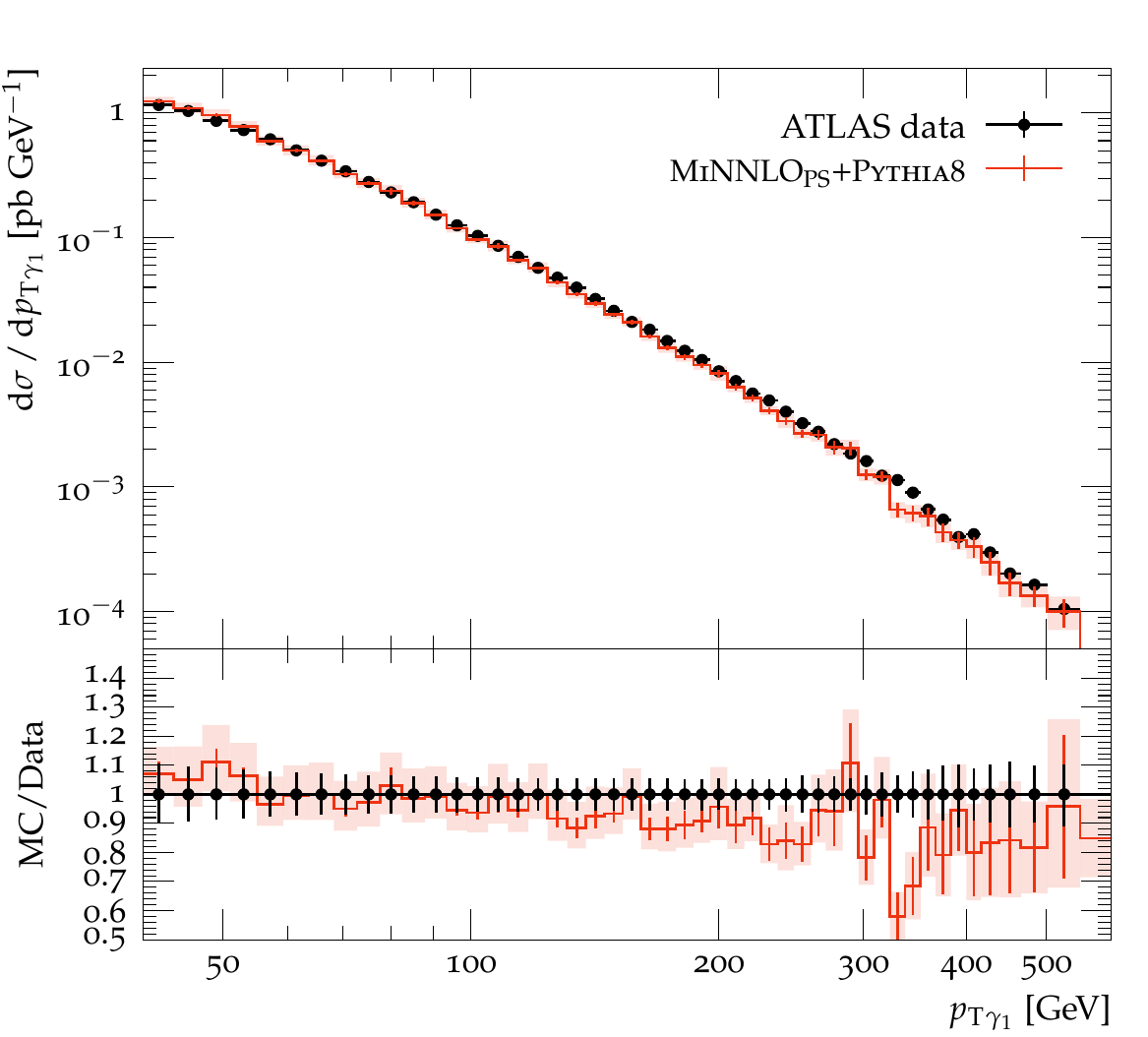}
    \includegraphics[width=0.49\textwidth]{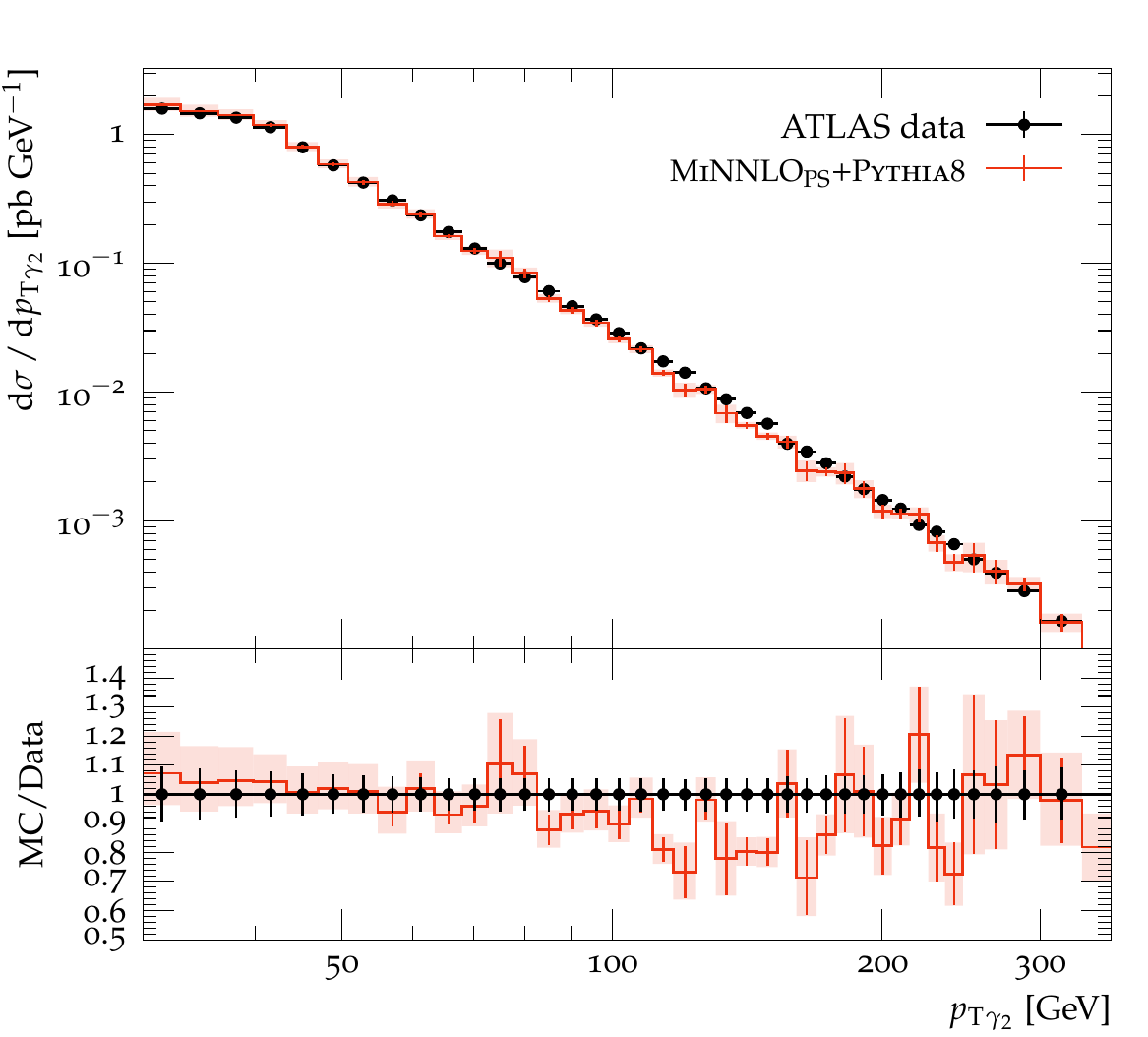}
  \end{center}
  \caption{Comparison between the ATLAS data and the distributions obtained
    with \MiNNLOPS{}+\PythiaEight{}, for the transverse momentum of the
    leading and subleading photon. Scale-variation bands are also shown,
    together with the statistical errors of the central value.}
  \label{fig:ATLAS_y_pts}
\end{figure}

In figures~\ref{fig:ATLAS_yy_m_pt} and~\ref{fig:ATLAS_y_pts}
we show the invariant mass and the transverse momentum of the diphoton
system, and the transverse momentum of the leading and subleading
photon, respectively. Overall we find a reasonably good agreement
between data and theoretical predictions.

For the invariant mass distribution $m_{\gamma\gamma}$ we observe a good
agreement in the bulk of the cross section. Given the cuts of
eq.~(\ref{eq:atlas_cuts}), the region where $m_{\gamma\gamma}< 80$~GeV
is populated only by $\gamma\gamma$ + jet(s) events, therefore
our results are only NLO+PS accurate, as confirmed also by the wider
uncertainty bands. For $m_{\gamma\gamma}\lesssim 40$~GeV, \MiNNLOPS{}
results overshoot ATLAS data by an amount compatible with what has been
observed, for other predictions of similar accuracy, in
ref.~\cite{ATLAS:2021mbt}. This is particularly true for the first bin.
The fact that this region is characterized
by a large NLO $K$-factor~\cite{Catani:2018krb} hints at the possibility
that the inclusion of higher-order corrections will improve the
agreement with data.  At large $m_{\gamma\gamma}$ values we observe
differences up to about 15\%. Such differences might be due to the
absence of higher-order effects. Top-quark mass effects
above the threshold $m_{\gamma\gamma}\simeq 2 m_t$, that we are
neglecting in the quark-induced 2-loop amplitudes as well as in the
$gg\to\gamma\gamma$ channel, can also induce differences at a
few percent level.

As far as the transverse-momentum spectrum of the diphoton pair is concerned,
our predictions are compatible with experimental data across most of the
$\ptAA$ spectrum, within the estimated theoretical uncertainties.
Due to the all-order treatment of soft and collinear radiation encoded in the
\MiNNLOPS{} Sudakov form factor and in the subsequent matching to \pythia{},
the region of small $\ptAA$ is in a fair agreement with the data, within the
statistical error and the scale-variation band. 
For large values of $\ptAA$, data and
theoretical predictions agree well in shape but display a flat offset of
about 10\%.

The comparison of our results against ATLAS data for the $\pt$ spectrum of
the leading and subleading photon is shown in
figure~\ref{fig:ATLAS_y_pts}. Data and theoretical predictions agree quite
well over the full range of the transverse momentum.

We would also like to point out that the choices for the central value of the
renormalization and factorization scales, in a fixed-order computation, can
have a sizable impact on the theory predictions at large invariant mass and
transverse momentum of the diphoton system, as discussed in detail e.g.~in
ref.~\cite{Gehrmann:2020oec}. The size of these effects are a consequence of
a slowly convergent perturbative series.
Assessing this type of effects by exploring different scale choices in our
\MiNNLOPS{} simulation goes beyond the scope of this paper, although this is
a-priori possible, both for the handling of the large-$\pt$ limit of the
singular component (first term of eq.~(\ref{eq:dsig_final})), as discussed
for instance in ref.~\cite{Mazzitelli:2021mmm}, as well as for the scales
used in the non-singular contribution (last term in the same equation).

\begin{figure}[htb!]
  \begin{center}  
    \includegraphics[width=0.49\textwidth]{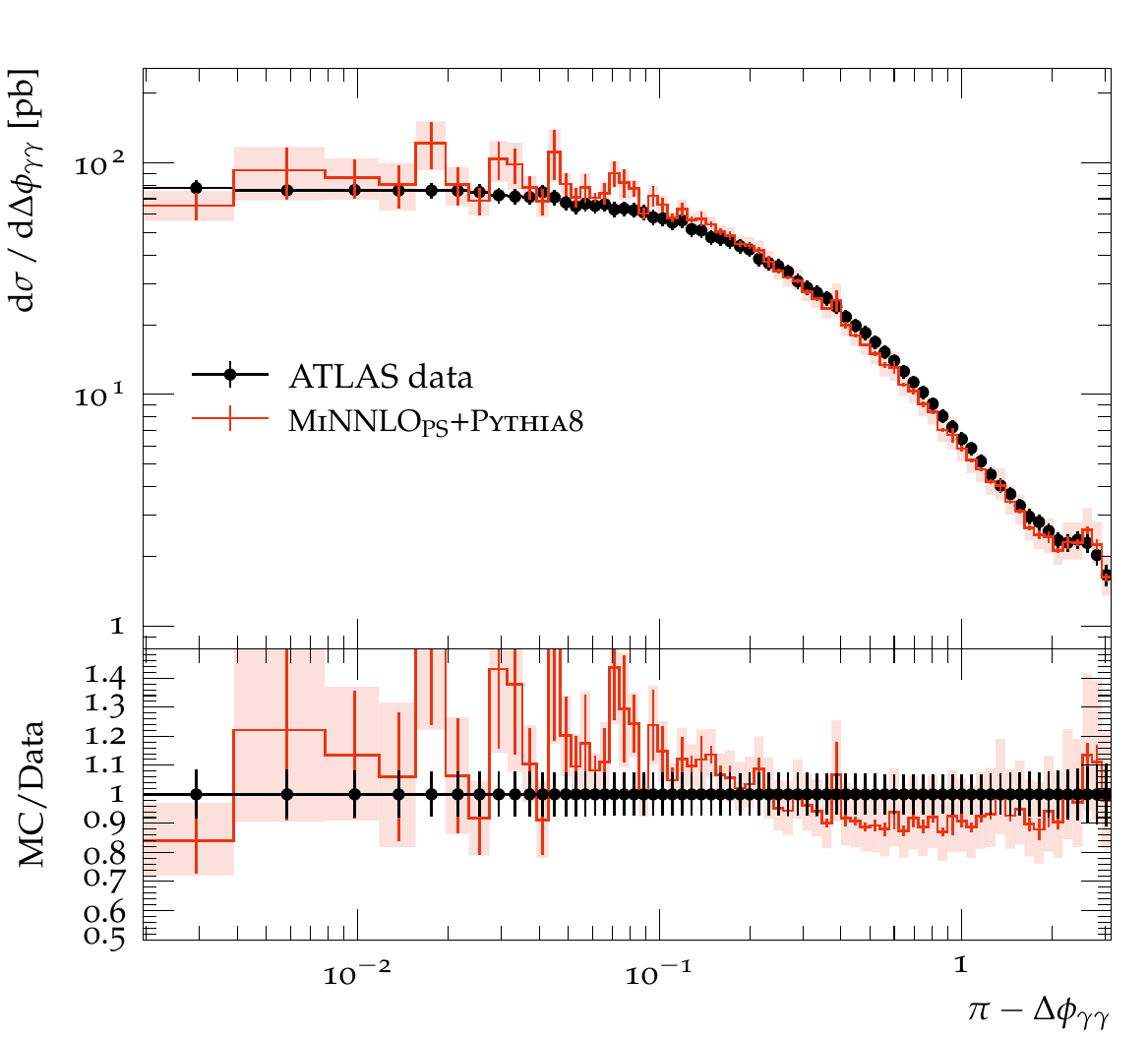}
    \includegraphics[width=0.49\textwidth]{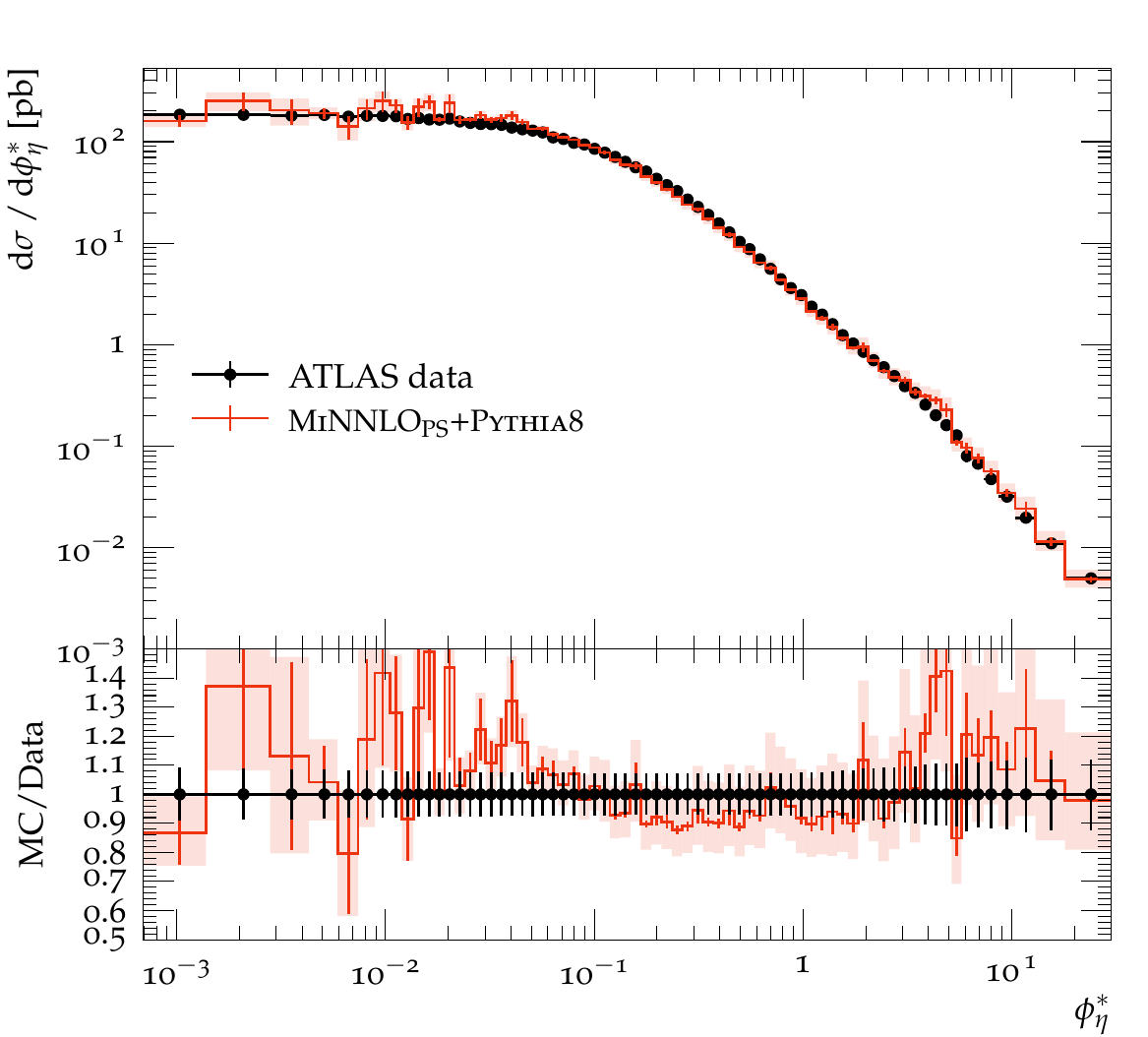}
  \end{center}
  \caption{Comparison between the ATLAS data and the distributions obtained
    with \MiNNLOPS{}+\PythiaEight{}, for the azimuthal separation of the two
    photons $\Delta\phi_{\gamma\gamma}$ and for \yyphiStar, as defined in
    eq.~(\ref{eq:phistar}).  Scale-variation bands are also shown, together
    with the statistical errors of the central value.}
  \label{fig:ATLAS_yy_phi}
\end{figure}

In figure~\ref{fig:ATLAS_yy_phi} we show the differential cross section as a
function of two angular observables: the azimuthal separation
$\Delta\phi_{\gamma\gamma}$ of the two photons and \yyphiStar, defined as
\begin{equation}
  \label{eq:phistar}
  \yyphiStar=\tan\frac{\yypiMDphi}{2}\,\sin\theta_{\eta}^* = \tan\frac{\yypiMDphi}{2}\,
  \sqrt{1-\left(\tanh\frac{\Delta y_{\gamma\gamma}}{2}\right)^2},
\end{equation}
where $\Delta y_{\gamma\gamma}$ is the azimuthal angular separation of the
two photons. Such variable, first introduced for Drell-Yan processes in
ref.~\cite{Banfi:2010cf}, is known to be sensitive to the same dynamics
governing the $\yypT$ spectrum, but it allows for a better resolution at
small values of $\yypT$.  The agreement between data and the theoretical
predictions is rather good on the whole range within the statistical and
scale-variation bands. We stress that, for $\Delta \phi_{\gamma\gamma} \to
\pi$, the NNLO fixed-order results for the above distributions are divergent,
while we get a finite distributions due to the {\sc\small MiNNLO} Sudakov
form factor.

\section{Conclusions}
\label{sec:conclusions}

In this paper we have presented the implementation of a new \MiNNLOPS{} event
generator accurate at NNLO in QCD for the production of a photon pair in
hadronic collisions, within the \pwgboxres{} framework.
This implementation is based on the \MiNNLOPS{} formalism presented in
refs.~\cite{Monni:2019whf, Monni:2020nks}, that we have modified in order to
deal with the new features needed to describe a process with photons at the
Born level.
We have devised and implemented a general method to deal with the presence of
QED singularities at the Born level within the \pwgboxres{}, and we have
presented a new mapping from $\gamma\gamma j$ to the $\gamma\gamma$.
The combined use of the new mapping and of the method to deal with the QED
divergences allowed us to simulate diphoton production at NNLO+PS accuracy
without introducing any generation or technical cut.
Our generator produces weighted events that cover the entire phase space,
thereby allowing the generated sample to be used with arbitrary fiducial
cuts.

Furthermore, we have introduced a few modifications to the \MiNNLOPS{}
formalism, that do not change the formal accuracy of the method and reduce
the numerical impact of higher-order terms, thus leading to better agreement
with fixed-order computations.  We have also verified that the new
formulation of \MiNNLOPS{}, when applied to the original \MiNNLOPS{}
implementations with massive bosons, gives results fully compatible with the
original one.

Finally, we have presented some distributions of physical interest obtained
from the generated events both before and after passing them through the
parton shower,
and compared them with the ATLAS
results at 13~TeV. Using the standard settings of \PythiaEight{} when
interfacing it with the \pwgboxres{}, we observe a reduction of about 5–10\%
of the inclusive differential cross sections compared to the
NNLO fixed-order ones.  We ascribe this effect to the increased
multiplicity of the partonic/hadronic activity coming from the shower causing
the photons to less likely satisfy the isolation criteria.  Further
investigation of the matching procedure between
\pwgboxres{}+\MiNNLOPS{} and \PythiaEight{} is left for future studies.

\acknowledgments
We would like to thank S.~Alioli, A.~Broggio and J.~Lindert for useful
discussions.
We also thank J.P.~Guillet, P.F.~Monni, and M.~Wiesemann
for carefully reading the manuscript and for their useful comments.

E.R. thanks P.F.~Monni and M.~Wiesemann for several discussions about
theoretical aspects related to the \MiNNLOPS{} method. In particular, the
relevant aspects of the formulation of the \MiNNLOPS{} method with the scale
choice discussed in section~\ref{sec:MiNNLO}, and their initial
implementation in the code, were developed in collaboration with them.

The work of A.G. was supported by the ERC Starting Grant REINVENT-714788.

\appendix

\section{The $\boldsymbol{\PhiBj} \to \boldsymbol{\PhiB}$ phase space
  projection}
\label{app:projection}

In section~\ref{sec:mapping} it was shown that one needs a mapping from the
$\PhiBj$ phase space to the $\PhiB$ one, and this mapping should be smooth
when $\pt \to 0$, i.e.~when the final-state jet is soft or collinear with the
incoming beams.

This projection plays a more important role for the process under study than
in previous processes dealt within the \MiNNLOPS{} procedure. In fact, for
diphoton production, we would like to avoid that the kinematics of a
$\gamma\gamma j$ event that is away from any collinear divergence is
projected on a kinematics for $\gamma\gamma$ production, where the photons
are quasi-collinear with the incoming beams, which would give divergent
contributions for the $pp \to \gamma\gamma$ matrix elements needed in the
\MiNNLOPS{} $D$ terms (see section~\ref{sec:MiNNLO}).  This projection is
determined by the requirements that it preserves the mass and rapidity of the
color singlet, and the direction of one of the photons in the laboratory
frame.

As far as the \MiNNLOPS{} formulae are concerned, we just need to define a
$\PhiBj \to \PhiB$ mapping that projects a $\PhiBj$ phase-space point to
a $\PhiB$ one.  In addition to this, we also provide the inverse mapping,
that will be used to compute the Jacobian of the transformation in
section~\ref{sec:jacobian}.

\subsection[The $\PhiBj$ kinematics]{The $\boldsymbol{\PhiBj}$ kinematics}
In the partonic center-of-mass frame, momentum conservation reads
 \begin{equation}
\label{eq:mom_conserv}
p_{\sss \oplus}+p_{\sss \ominus} = \pgo + \pgt +\pj\,,
\end{equation}
where 
\begin{equation}
p_{\sss \oplus} = \frac{\sqrt{s}}{2}(1,0,0,1)\,, \qquad \qquad
p_{\sss \ominus} = \frac{\sqrt{s}}{2}(1,0,0,-1)\,,
\end{equation}
and $s$ is the squared center-of-mass energy.  In this section, at difference
with the rest of the paper, we denote with $\pgo$ and $\pgt$ the momenta of
the two photons, irrespectively of which  one is the hardest or the
softest.
Using the standard FKS~\cite{Frixione:1995ms} radiation variables $\xi$, $y$
and $\phij$, we write the jet momentum as
\begin{equation}
\label{eq:pj}  
  \pj = \frac{\sqrt{s}}{2} \,  \xi \(
    1, \,
    \sin\thj\sin\phij, \,
    \sin\thj\cos\phij, \,
    \cos\thj
  \), 
\end{equation}
where $\cos\thj = y$.
Using eq.~(\ref{eq:mom_conserv}), the momentum of the diphoton system is then
given by
\begin{equation}
  \label{eq:pgg_CMggj}
  p_{\gamma\gamma} = \frac{\sqrt{s}}{2} \, \(
  2-\xi, \,
    -\xi \sqrt{1-y^2}\sin\phij, \,
    -\xi \sqrt{1-y^2}\cos\phij, \,
    - \xi y
  \),
\end{equation}
that has invariant mass and rapidity given by
\begin{equation}
  \label{eq:y_gg^CM}
  m_{\gamma\gamma} = \sqrt{s \( 1-\xi \)}\,,
\qquad \qquad
  y_{\gamma\gamma}^{\sss\rm CM} = \frac{1}{2} \log\! \( \frac{2-\xi-\xi y}{2-\xi+\xi y} \)\,.
\end{equation}

\subsection[From $\PhiB$ to $\PhiBj$]{From $\boldsymbol{\PhiB}$ to $\boldsymbol{\PhiBj}$}
In order to preserve the direction of one photon, we need to express the
momenta of the involved particles in two frames where the diphoton system has
the same rapidity.
We choose to work in the center-of-mass frame of the $\aaj$ system
because, in this frame, the momentum of the final-state parton has a simple
representation in terms of the FKS variables, as given in eq.~(\ref{eq:pj}).
We parametrize the momenta of the two photons in their center-of-mass frame as
\begin{equation}
  \lg
  \begin{array}{l}
    \displaystyle
    \bar{p}_{\gamma_1} = \frac{m_{\gamma\gamma}}{2} \(
    1 ,\,\sin\thb\sin\phib,\, \sin\thb\cos\phib,\,\cos\thb \),
    ~\\[3mm]
    \displaystyle
    \bar{p}_{\gamma_2} = \frac{m_{\gamma\gamma}}{2} \( 1,\,
    -\sin\thb\sin\phib ,\,    -\sin\thb\cos\phib ,\,  -\cos\thb \),
  \end{array}
  \right.
\end{equation}
and we then perform a longitudinal boost with rapidity $y_{\gamma\gamma}^{\sss\rm
  CM}$ of eq.~(\ref{eq:y_gg^CM}) to obtain
\begin{equation}
  \lg
  \begin{array}{l}
    \displaystyle
    \ppgo  =   \frac{m_{\gamma\gamma}}{2} \(
  \frac{2-\xi-\xi y\cos\thb}{r} , \,
    \sin\thb\sin\phib, \,
    \sin\thb\cos\phib, \,
    \frac{(2-\xi) \cos\thb - \xi y}{r}  \),
     \\[3mm]
    \displaystyle
    \ppgt  =   \frac{m_{\gamma\gamma}}{2} \(
  \frac{ 2-\xi+\xi y\cos\thb}{r}, \,
    -\sin\thb\sin\phib, \,
    -\sin\thb\cos\phib, \,
    \frac{-(2-\xi) \cos\thb - \xi y }{r}  \),
  \end{array}
  \right.
\end{equation}
where
\begin{equation}
r =  \sqrt{(2-\xi)^2 - \xi^2 y^2}\,.
\end{equation}
We now impose that the momentum $\pgo$ in the partonic
center-of-mass frame of~$\aaj$ has the same direction as $\ppgo$.
In this way, if the photon $\gamma_1$ of $\PhiB$ becomes too close to the
incoming beams, then also the photon $\gamma_1$ of $\PhiBj$ will be close to the
beams, and will be suppressed by the Born suppression factors of
eq.~(\ref{eq:supp_fact_Born_MiNNLO}). To achieve this, we
  change the energy of $\ppgo$, introducing the dimensionless parameter
  ${\cal E}$, such that the momentum of $\gamma_1$ in
  the center-of-mass frame  kinematics of $\PhiBj$ is given by
\begin{equation}
  \label{eq:pg1}
    \pgo  =  {\cal E} \, \sqrt{s}  \(
  \frac{2-\xi-\xi y\cos\thb}{r} , \,
    \sin\thb\sin\phib, \,
    \sin\thb\cos\phib, \,
    \frac{(2-\xi) \cos\thb - \xi y}{r}  \).
\end{equation}
We fix the value of ${\cal E}$ by imposing that the second photon stays
massless. In fact, from the momentum conservation of eq.~(\ref{eq:mom_conserv}) 
\begin{eqnarray}
\label{eq:pg2}
\pgt  &=& p_{\sss \oplus}+p_{\sss \ominus} -\pgo  -\pj
\nonumber\\
&=& \sqrt{s}
\(
1 -\frac{\xi}{2}-{\cal E}\,\frac{2-\xi-\xi y\cos\thb}{r},\,
-\frac{\xi}{2}\sqrt{1-y^2}\sin\phij -{\cal E}\sin\thb\sin\phib,\,
\right.
\nonumber\\
 && \hspace{8mm}\left.
-\frac{\xi}{2}\sqrt{1-y^2}  \cos\phij - {\cal E}\sin\thb\cos\phib,\,
-\frac{\xi y}{2}  - {\cal E}\,\frac{(2-\xi) \cos\thb - \xi y}{r}
\),
\end{eqnarray}
and by imposing $\pgt^2  = 0$ we have
\begin{equation}
    {\cal E}  = \frac{ 1-\xi }{r +
    \xi\sqrt{1-y^2} \sin \thb \cos(\phij-\phib)}\,.
\end{equation}
Once the momenta $\pj$, $\pgo$ and $\pgt$, given by eqs.~(\ref{eq:pj}),
(\ref{eq:pg1}) and~(\ref{eq:pg2}) are known in their partonic center-of-mass
frame, we can obtain their expression in the laboratory system by performing
a longitudinal boost with the rapidity of the center-of-mass system. 

\subsection{The Jacobian of the mapping}
\label{sec:jacobian}
In order to conclude the procedure to build $\PhiBj$ from $\PhiB$, we need to
compute the Jacobian of the transformation we have outlined in the previous
section.  Following the \Powheg{} notation~\cite{Frixione:2007vw, Alioli:2010xd},
we write the infinitesimal phase space volume in terms of the Bjorken $x$ as
\begin{equation}
\mathd\Phi_{3} \equiv \mathd x_{\sss \oplus} \, \mathd x_{\sss \ominus} \,\mathd\Phi_{\gamma\gamma j} = {\mathd \bar{x}_{\sss \oplus} \, \mathd \bar{x}_{\sss \ominus}}
\,\mathd\Phirad \,\mathd\Phi_{\gamma\gamma}\,,
\end{equation}
where
\begin{equation}
\mathd\Phirad \equiv \frac{s}{(4\pi)^3}\, \frac{\xi}{1-\xi}\,\mathd \xi\, \mathd y \,\mathd \phij\,,
\end{equation}
is written using the FKS radiation variables.
We would like to express  $\mathd\Phi_{\gamma\gamma}$  using the 
polar and azimuthal angle of one photon in the center-of-mass frame of
the diphoton system, i.e.~$\thp$ and $\phip$, where it can
be written as
\begin{equation}
 \mathd\Phi_{\gamma\gamma} = \frac{1}{32 \pi^2}\, \mathd \cos\thp \, \mathd \phip\,,
\end{equation}
in terms of the angles of our mapping, i.e.~$\thb$ and $\phib$, that we have
used to write eqs.~(\ref{eq:pg1}) and~(\ref{eq:pg2}). Using the fact that
$\mathd\Phi_{\gamma\gamma}$ is a Lorentz scalar, we need to compute the
Jacobian $J$ that relates the change of variables
\begin{equation}
\mathd\Phi_{\gamma\gamma} = \frac{J}{32 \pi^2} \,\mathd \cos\thb \, \mathd
\phib \,.
\end{equation}

In order to do this, we find the sequence of boosts from the partonic
center-of-mass frame of $\gamma\gamma j$ to the center-of-mass frame of the
diphoton system.  We first perform a longitudinal boost of the momenta
$\pgo$ and $\pgt$ of eqs.~(\ref{eq:pg1}) and~(\ref{eq:pg2}) to the frame
where the diphoton system has zero rapidity.  The longitudinal rapidity of
the boost is given by eq.~(\ref{eq:y_gg^CM}) and we obtain
\begin{eqnarray}
    \displaystyle  
    \pgo^{\sss\rm L}  &=&  {\cal E} \, \sqrt{s}
    \(
    1 , \,
    \sin\thb\sin\phib, \,
    \sin\thb\cos\phib, \,
    \cos\thb  \),
\nonumber    
  \\[3mm]
  \displaystyle
    \pgt^{\sss\rm L}  &=&   \sqrt{s}
    \(
    \frac{r}{2}  - {\cal E},\,
    -\frac{\xi}{2}\sqrt{1-y^2}\sin\phij -{\cal E}\sin\thb\sin\phib,\,
    \right.
    \nonumber
    \\[3mm]
    \displaystyle
    && \mbox{}\hspace{20mm}
    \left.
    -\frac{\xi}{2}\sqrt{1-y^2}  \cos\phij - {\cal E}\sin\thb\cos\phib,\,    
    -{\cal E}  \cos\thb  \).
\end{eqnarray}
Then we perform a transverse boost to the diphoton rest system with
transverse velocity
\begin{equation}
\boldsymbol{v_{\sss\rm T}} = -\frac{\xi\sqrt{1-y^2}}{r}\(\sin\phij, \,
\cos\phij,\, 0 \),
\end{equation}
and we get
\begin{equation}
  \pgo^{\sss\rm T}  =  \frac{m_{\gamma\gamma}}{2}
  \( 1,   \boldsymbol{\pgo^{\sss\rm T}} \),  \qquad \qquad
  \pgt^{\sss\rm T}  =  \frac{m_{\gamma\gamma}}{2}
  \( 1,  - \boldsymbol{\pgo^{\sss\rm  T}} \),
\end{equation}    
where
\begin{eqnarray}
  \label{eq:pg1T_bar}
  \boldsymbol{\pgo^{\sss\rm T}}
  &=& \frac{1}{r +
    \xi\sqrt{1-y^2} \sin \thb \cos(\phij-\phib)} 
  \nonumber\\
  && \times 
  \Bigg(
    \sin \phij \left(r \sin \thb \cos
  (\phib-\phij)+\xi  \sqrt{1-y^2}\right)
  +2
    \sqrt{1-\xi} \cos \phij \sin \thb \sin
    (\phib-\phij)\,,
    \nonumber
    \\[1mm]
    && \hspace{6mm}
      \cos \phij \left(r \sin \thb \cos
    (\phib-\phij)+\xi  \sqrt{1-y^2}\right)- 2 \sqrt{1-\xi } \sin \phij \sin \thb \sin
    (\phib-\phij)\,,
\nonumber    
\\[1mm]
&&  \hspace{6mm}  2  \sqrt{1-\xi }\cos \thb  \Bigg).
\end{eqnarray}
In the diphoton rest system, $\boldsymbol{\pgo^{\sss\rm T}}$, written as a
function of $\thp$ and $\phip$, has components
\begin{equation}
\label{eq:pg1T_prime}
  \boldsymbol{\pgo^{\sss\rm T}}
  \equiv
   \( \sin\thp\sin\phip,\,
    \sin\thp\cos\phip, \,
    \cos\thp
  \).
\end{equation}
By comparing eqs.~(\ref{eq:pg1T_bar}) and~(\ref{eq:pg1T_prime}), we can write
\begin{eqnarray}
  \cos\thp &=&  \frac{ 2  \sqrt{1-\xi }\cos \thb}{r +
    \xi\sqrt{1-y^2} \sin \thb \cos(\phij-\phib)} 
  \\[2mm]
\tan\phip &= &\frac{ \sin \phij \left(r \sin \thb \cos
  (\phib-\phij)+\xi  \sqrt{1-y^2}\right)
  +2
    \sqrt{1-\xi} \cos \phij \sin \thb \sin
    (\phib-\phij) }{\cos \phij \left(r \sin \thb \cos
    (\phib-\phij)+\xi  \sqrt{1-y^2}\right)- 2 \sqrt{1-\xi } \sin \phij \sin \thb \sin
  (\phib-\phij)}
\nonumber\\
  \end{eqnarray}
and from these expressions, we can compute the Jacobian $J$
  \begin{equation}
  J = \frac{\partial{\( \cos\thp, \phip\)}}{\partial\(\cos\thb, \phib \)}=
  \frac{4 (1-\xi)}
       {\left( r + \xi\sqrt{1-y^2} \sin\thb \cos (\phib-\phij)\right)^2}\,.
  \end{equation}
This completes the calculation of the direct and inverse mapping $\PhiBj
\leftrightarrow \PhiB$ and the construction of the three-body phase space,
from one side, and of its inverse mapping, on the other.

\section{\MiNNLOPS{} formulae}
\label{app:scale_dep}
In this section, we provide the explicit expressions for the ingredients
needed in the \MiNNLOPS{} formulae, focussing in particular on their
renormalization and factorization scale dependence.  We use the same notation
as ref.~\cite{Monni:2019whf}, and we refer the interested reader to this
paper for further details.
For ease of notation, we drop all the dependences on the kinematic variables,
and we retain only that on the scales, which we assume to be equal,
$\muR=\muF=\mu$.  In section~\ref{sec:MiNNLO}, a physical quantity $G$
computed at scale $\mu$ was indicated with $\lq G\!\(\PhiB,\pt\) \rq_\mu$. In
the compact notation of this section, is is now simply written as $G(\mu)$.

Using this notation, from the definition of $D$ in eq.~(\ref{eq:Dterms}) and
of its expansion in eq.~(\ref{eq:D_expans_pt}), we have
\begin{eqnarray}
  \lq D \!\(\pt \) \rq^{(1)} &=&
  \lq\frac{d{\Lum} \!\(\pt \)}{d\pt}\rq^{\!(1)} -
  \lq\Lum \!\(\pt\)\rq^{(0)}
  \lq\frac{d\tilde{S} \!\(\pt\)}{d\pt}\rq^{\!(1)}
  \\
  \lq D \!\(\pt\) \rq^{(2)} &=&
  \lq\frac{d{\Lum} \!\(\pt\)}{d\pt}\rq^{\!(2)} -
  \lq\Lum \!\(\pt\)\rq^{(0)}
  \lq\frac{d\tilde{S} \!\(\pt\)}{d\pt}\rq^{\!(2)} - \lq\Lum \!\(\pt\)\rq^{(1)}
  \lq\frac{d\tilde{S} \!\(\pt\)}{d\pt}\rq^{\!(1)}.
\end{eqnarray}
The explicit expression for the luminosity $\Lum$, evaluated at scales $\pt$,
is
\begin{equation}
  \label{eq:lum_def}
  \Lum\!\(\pt\) = \sum_{cd} \left|M_{cd\to\gamma\gamma}\right|^2 \tilde{H}_{cd\to\gamma\gamma}
  \sum_{ij} \(\tilde{C}_{ci} \otimes f_i^{[a]} \)
  \(\tilde{C}_{dj} \otimes f_j^{[b]} \),
\end{equation}
where $\left|M_{cd\to\gamma\gamma}\right|^2$ is the Born squared amplitude
for the process $c d \,\to\, \gamma \gamma$ (in this case $c$ and $d$ are a
quark-antiquark pair), $\tilde{H}_{cd \to \gamma\gamma}$ encodes its virtual
corrections up to two loops,\footnote{In the two-loop contributions, all
effects from massive quarks have been neglected.} the $\tilde{C}_{ij}$ are
the collinear coefficient functions and $f_i^{[a]}$ is the PDF for the parton
$i$ in the hadron $a$.
 Both $\tilde{H}_{cd \to \gamma\gamma}$ and $\tilde{C}_{ij}$ can be written
 as an expansion in $\as$ as
\begin{equation}
  \tilde{H}_{cd\to\gamma\gamma} = 1 + \(\frac{\as}{2\pi}\) H^{(1)} +
  \(\frac{\as}{2\pi}\)^2 \tilde{H}^{(2)} + \ord{\as^3},
\end{equation}
and
\begin{equation}
  \tilde{C}_{ij}(z) = \delta (1-z) \, \delta_{ij} +
  \(\frac{\as}{2\pi}\) C_{ij}^{(1)}(z) + \(\frac{\as}{2\pi}\)^2
  \tilde{C}_{ij}^{(2)}(z) + \ord{\as^3}.
\end{equation}
In the above formula $H^{(1)}$ and $\tilde{H}^{(2)}$ are the one- and
two-loop virtual corrections for the process $q \bar{q} \to \gamma \gamma$
written in the \MSB{} renormalization and subtraction schemes.\footnote{The
$\tilde{H}^{(2)}$ term is obtained from the two-loop contribution $H^{(2)}$
of refs.~\cite{Anastasiou:2002zn, Catani:2013tia}, by the manipulations
required by the \MiNNLOPS{} method, as explained in
ref.~\cite{Monni:2019whf}.}
If we introduce the usual definitions for the Mandelstam variables
\begin{equation}
  s = \(p_{\sss\oplus}+p_{\sss\ominus}\)^2 \qquad t =
  \(p_{\sss\oplus}-p_{\gamma_1}\)^2 \qquad u =
  \(p_{\sss\oplus}-p_{\gamma_2}\)^2,
\end{equation}
and define
\begin{equation}
  v = -\frac{u}{s},
\end{equation}
the explicit expression for $H^{(1)}$ is~\cite{Aurenche:1985yk}
\begin{eqnarray}
  H^{(1)} &=& \CF \left\{\pi^2 - 7 + \frac{1+\(1-v\)^2}{v^2+\(1-v\)^2}
  \, \log^2\!\(1-v\) + \frac{1+v^2}{v^2+\(1-v\)^2} \, \log^2\!\(v\)
  \right.
  \nonumber \\
  && \left. {} + \frac{v\(2+v\)}{v^2+\(1-v\)^2} \, \log\!\(1-v\) +
  \frac{\(1-v\)\left[2+\(1-v\)\right]}{v^2+\(1-v\)^2} \, \log\!\(v\)
  \right\}.
\end{eqnarray}
We stress again that
the renormalization and factorization scales in eq.~(\ref{eq:lum_def})
are both set to $\pt$, as indicated by the argument of the luminosity
$\Lum$.
Finally, the convolution operator is defined, as usual, as
\begin{equation}
  \(f \otimes g\)\!\(x\) \equiv \int_x^1 \frac{dz}{z} f\!\(z\)
  g\!\(\frac{x}{z}\).
\end{equation}
The expansion of $D$ at a scale $\mu$ can be written as
\begin{equation}
\label{eq:D_expans_mu}
D\!\(\mu\) =
\frac{\as(\mu)}{2\pi} \lq D\!\(\mu\) \rq^{(1)} +
\(\frac{\as(\mu)}{2\pi}\)^2 \lq D\!\(\mu\) \rq^{(2)} + \ord{\as^3}
\end{equation}
where
\begin{eqnarray}
  \label{eq:D1_mu_anal}
  \lq D \!\(\mu\) \rq^{(1)} &=&
  \lq\frac{d{\Lum} \!\(\mu\)}{d\pt}\rq^{\!(1)} -
  \lq\Lum \!\(\mu\)\rq^{(0)}
  \lq\frac{d\tilde{S} \!\(\mu\)}{d\pt}\rq^{\!(1)}
  \\
 \label{eq:D2_mu_anal}
  \lq D \!\(\mu\) \rq^{(2)} &=&
  \lq\frac{d{\Lum} \!\(\mu\)}{d\pt}\rq^{\!(2)} -
  \lq\Lum \!\(\mu\)\rq^{(0)}
  \lq\frac{d\tilde{S} \!\(\mu\)}{d\pt}\rq^{\!(2)} - \lq\Lum \!\(\mu\)\rq^{(1)}
  \lq\frac{d\tilde{S} \!\(\mu\)}{d\pt}\rq^{\!(1)}.
\end{eqnarray}
All the needed ingredients to compute eqs.~(\ref{eq:D1_mu_anal})
and~(\ref{eq:D2_mu_anal}) can be obtained by applying the DGLAP evolution equations
\begin{equation}
  \frac{d}{d\log\muf^2} \,f_c^{[a]}\!\(\muf\) =
  \frac{\as}{2\pi} \sum_i \(P_{ci}^{(0)} \otimes f_i^{[a]}\)
  + \(\frac{\as}{2\pi}\)^2
  \sum_i \(P_{ci}^{(1)} \otimes f_i^{[a]}\) + \ord{\as^3},
\end{equation}
to compute the scale dependence of the parton distribution function
\begin{equation}
\label{eq:RGE_pdf}
f_c^{[a]}\!\(\muf\) = f_c^{[a]}\!\(\pt\) + \frac{\as}{2\pi}
\sum_i \(P_{ci}^{(0)} \otimes f_i^{[a]}\)
\log\frac{\muf^2}{\pt^2} + \ord{\as^2},
\end{equation}
and the renormalization group equation 
\begin{equation}
  \frac{d}{d\log\mur^2} \frac{\as\(\mur\)}{2\pi} = - \(\frac{\as}{2\pi}\)^2 \beta_0 + \ord{\as^2}
\end{equation}
with
\begin{equation}
  \beta_0 = \frac{11\CA-4\TR\nf}{6},
\end{equation}
to compute the running of $\as$
\begin{equation}
\label{eq:RGE_as}
\frac{\as\(\mur\)}{2\pi} = \frac{\as\(\pt\)}{2\pi} - \(\frac{\as}{2\pi}\)^2 \beta_0  \log\frac{\mur^2}{\pt^2}
+\ord{\as^3}.
\end{equation}
We then  obtain
\begin{eqnarray}
   \lq\Lum\(\mu\)\rq^{(0)}  &=&  \sum_{cd}
  \left|M_{cd\to\gamma\gamma}\right|^2 f_c^{[a]} f_d^{[b]}
  \\
   \lq\Lum\(\mu\)\rq^{(1)}  &=&  \sum_{cd}
  \left|M_{cd\to\gamma\gamma}\right|^2 \Bigg\{ \sum_i \lq \(C_{ci}^{(1)}
    \otimes f_i^{[a]}\) f_d^{[b]} + f_c^{[a]} \(C_{di}^{(1)}
    \otimes f_i^{[b]}\) \rq
  \nonumber \\
  &&{} -  \sum_i \lq \(P_{ci}^{(0)} \otimes f_i^{[a]}\)
    f_d^{[b]} + f_c^{[a]} \(P_{di}^{(0)} \otimes f_i^{[b]}\) \rq
  \log\!\(\frac{\muf^2}{\pt^2}\)
  \nonumber \\
  && {} + H_{cd\to\gamma\gamma}^{(1)}\, f_c^{[a]} f_d^{[b]}
  \Bigg\},
\end{eqnarray}
where the parton distribution functions  on the right-hand side are evaluated at scale
$\muf$.

In addition, from the corresponding expressions evaluated at scale $\pt$
given in refs.~\cite{Monni:2019whf, Monni:2020nks}, we can compute
\begin{eqnarray}
  \lq\frac{d\Lum(\mu)}{d\pt}\rq^{\!(1)}  &=&  \sum_{cd}
  \left|M_{cd\to\gamma\gamma}\right|^2 \sum_i \lq\(P_{ci}^{(0)} \otimes
    f_i^{[a]}\) f_d^{[b]} + f_c^{[a]} \(P_{di}^{(0)} \otimes
    f_i^{[b]}\)\rq
  \\
  \lq\frac{d\Lum(\mu)}{d\pt}\rq^{\!(2)}  &=&  \sum_{cd}
  \left|M_{cd\to\gamma\gamma}\right|^2 \Bigg\{ \sum_i \lq \(P_{ci}^{(1)}
    \otimes f_i^{[a]}\) f_d^{[b]} + f_c^{[a]} \(P_{di}^{(1)} \otimes
    f_i^{[b]}\) \rq
  \nonumber \\
  &&{}+  \sum_{ij} \bigg[ \(P_{ci}^{(0)} \otimes
    C_{ij}^{(1)} \otimes f_j^{[a]}\) f_d^{[b]} + \(C_{ci}^{(1)}
    \otimes f_i^{[a]}\) \(P_{dj}^{(0)} \otimes f_j^{[b]}\)
    \nonumber \\
 && {}\hspace{1.3cm} + \(P_{ci}^{(0)} \otimes f_i^{[a]}\)
    \(C_{dj}^{(1)} \otimes f_j^{[b]}\) + f_c^{[a]} \(P_{di}^{(0)}
    \otimes C_{ij}^{(1)} \otimes f_j^{[b]}\) \bigg]
  \nonumber \\
  && {} -  \sum_{ij} \bigg[ \(P_{ci}^{(0)} \otimes
    P_{ij}^{(0)} \otimes f_j^{[a]}\) f_d^{[b]} + \(P_{ci}^{(0)}
    \otimes f_i^{[a]}\) \(P_{dj}^{(0)} \otimes f_j^{[b]}\)
    \nonumber \\
    && {} \hspace{1.3cm} + \(P_{ci}^{(0)} \otimes f_i^{[a]}\)
    \(P_{dj}^{(0)} \otimes f_j^{[b]}\) + f_c^{[a]} \(P_{di}^{(0)}
    \otimes P_{ij}^{(0)} \otimes f_j^{[b]}\) \bigg]
  \log\!\(\frac{\muf^2}{\pt^2}\)
  \nonumber \\
  &&{}+  H_{cd\to\gamma\gamma}^{(1)} \sum_i \lq
    \(P_{ci}^{(0)} \otimes f_i^{[a]}\) f_d^{[b]} + f_c^{[a]}
    \(P_{di}^{(0)} \otimes f_i^{[b]}\)\rq \Bigg\}
  \nonumber \\
  && {}- \beta_0 \lq \Lum\(\mu\) \rq^{(1)}
  + \beta_0 \lq\frac{d\Lum\!\(\mu\)}{d\pt}\rq^{\!(1)} \log\!\(\frac{\mur^2}{\pt^2}\),
\end{eqnarray}
and
\begin{eqnarray}
  \lq\frac{d\tilde{S}\!\(\mu\)}{d\pt}\rq^{\!(1)}  &=&  A^{(1)}
  \log\!\(\frac{Q^2}{\pt^2}\) + B^{(1)}
  \\
  \lq\frac{d\tilde{S}\!\(\mu\)}{d\pt}\rq^{\!(2)}  &=&  A^{(2)}
  \log\!\(\frac{Q^2}{\pt^2}\) + \tilde{B}^{(2)} + \beta_0
  \lq\frac{d\tilde{S}\!\(\mu\)}{d\pt}\rq^{\!(1)}
  \log\!\(\frac{\mur^2}{\pt^2}\).
\end{eqnarray}
where ${A}^{(i)}$ and $B^{(i)}$ are given, for example, in
ref.~\cite{Monni:2019whf}.

\bibliography{paper}
    
\end{document}